\documentclass[%
 aps,
 prl,
 longbibliography,
 sd,%
 amsmath,amssymb,
 reprint,%
]{revtex4-1}

\usepackage[dvipdfmx]{graphicx}
\usepackage{dcolumn}
\usepackage{bm}

\usepackage{setspace,color,url}
\usepackage{comment}
\begin{document}

\preprint{Submitted to Physical Review **}

\title{Membranes for spontaneous separation of pedestrian counter flows}
%
\author{Shihori Koyama}
\affiliation{Toyota Central R\&D Labs., Inc., Bunkyo-ku, Tokyo 112-0004, Japan}
\author{Daisuke Inoue}
\affiliation{Toyota Central R\&D Labs., Inc., Bunkyo-ku, Tokyo 112-0004, Japan}
\author{Akihisa Okada}
\affiliation{Toyota Central R\&D Labs., Inc., Bunkyo-ku, Tokyo 112-0004, Japan}
\author{Hiroaki Yoshida}
\email{h-yoshida@mosk.tytlabs.co.jp}
\affiliation{Toyota Central R\&D Labs., Inc., Bunkyo-ku, Tokyo 112-0004, Japan}
\date{\today}
%
\begin{abstract}
Designing efficient traffic lanes for pedestrians is a critical aspect of urban planning as walking remains the most common form of mobility among the increasingly diverse methods of transportation. Herein, we investigate pedestrian counter flows in a straight corridor, in which two groups of people are walking in opposite directions. 
We demonstrate, using a molecular dynamics approach applying the social force model, that a simple array of obstacles improves flow rates by producing flow separations even in crowded situations. We also report on a developed model describing the separation behavior that regards an array of obstacles as a membrane and induces spontaneous separation of pedestrians groups.
When appropriately designed, those obstacles are fully capable of controlling the filtering direction so that pedestrians tend to keep moving to their left (or right) spontaneously. 
These results have the potential to provide useful guidelines for industrial designs aimed at improving ubiquitous human mobility. 
\end{abstract}
\maketitle
%
%
%
%
\section{\label{sec_intro}Introduction}

Modern transportation systems are becoming increasingly complex and often require different time and spatial scales, as represented by the rapid growth of the diverse transportation methods and mobility technologies~\cite{BBG+2018, Barthelemy2019}. 
Since there are still numerous phenomena that are not fully understood within each of such transportation systems, both experimental and theoretical studies aimed at understanding such phenomena have been performed continuously~\cite{LDK+2014,VKT+2016}. Among the different transportation methods, walking remains the most fundamental, so pedestrian flows have been widely studied~\cite{Scott1974,PSU1983,GM1985,VE1994}. In typical experimental studies, pedestrian trajectories are observed and analyzed by recording their motions with video cameras or using laser measurements~\cite{SSK+2005,HJA2007,JHA+2008, CSC2009,MHG+2009, ZIK2011,ZKS+2011, ZKS+2012}. 
On the other hand, theoretical approaches have also been used to gain a systematic understanding of observed pedestrian behaviors and/or for predicting the pedestrian flows under various circumstances~\cite{Helbing1992,BKS+2001,Bonabeau2002,Hughes2002,SSL2006}. 
For example, the so-called social force model, first proposed by Helbing and Moln\'ar~\cite{HM1995}, is one of the most widely used theoretical approaches used to model pedestrian movements, which enables us to simulate flows using molecular dynamics~\cite{HFV2000,LKF2005,OVH+2016,SCF+2017,KZL2019}.

\begin{figure}[t]
\label{fig:corrider}
	\begin{center}
	\includegraphics[width=0.95\hsize]{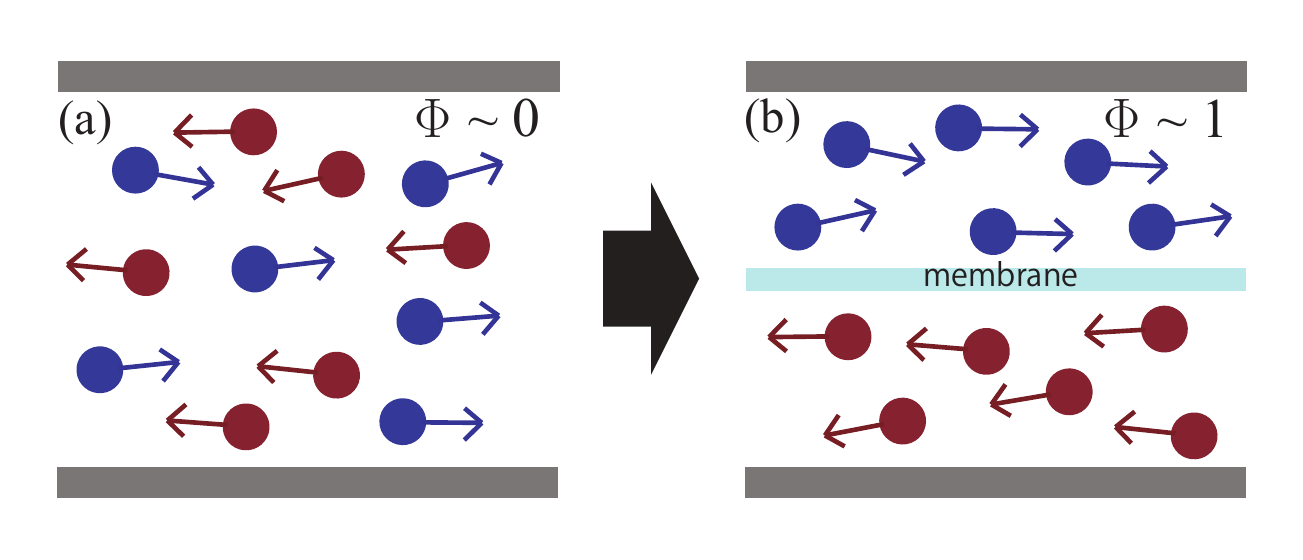}
	\caption{Pedestrian counter flow separation. (a) flow without filtering membrane and (b) flow separated into two streams using a permeable membrane, which is realized by an array of obstacles. 
The parameter $\Phi$ shown in each figure 
is an order parameter indicating the degree of lane-formation (or separation). See Eq.~\eqref{eq_phi} for the precise definition.
}
	\label{fig_cf}
	\end{center}
\end{figure}

In the present study, we also employ the social force model to investigate pedestrian behaviors in a straight corridor in which two groups of people are walking in opposite directions. 
Similar situations with two groups of particles moving in the other directions have been extensively
studied in the context of lane and pattern formations
not only of pedestrians~\cite{IK2017,FMN2018}
but also of various physical particles, such as charged colloids~\cite{VBI2011,VWR+2011,TEL2019}, microswimmers~\cite{KK2015}, and
plasmas~\cite{SWI+2009,SBG2020}. These studies are focused mainly on
a bulk system without obstacles.
Here we demonstrate that separation-membrane-like obstacles placed along the centerline of a corridor improve flow efficiency even in crowded situations. This is because the presence of those obstacles triggers a spontaneous separation of the pedestrians groups, thereby resulting in an unconscious ``keep-left'' pedestrian mentality, as illustrated in Fig.~\ref{fig_cf}. 
Although relevant studies have been reported, such as effects of placing columns asymmetrically near an exit~\cite{HFV2000,SSY2019},
and a partition line effect controlling the critical density in the jamming transition~\cite{TTN2002},
the enhancement of a pedestrian counter flow by means of particular choices of obstacles is, to our knowledge, new. 
We also report on the development of a model describing the separation behavior, which is inspired by a reminiscent membrane separating multi-component fluids. These findings related to pedestrian group filtering could potentially provide useful guidelines for improving daily pedestrian flows.

%
%
\section{\label{sec_problem}Problem}

Here we consider a throng of $N$ pedestrians walking in a straight corridor of width $W$ and length $L$, as shown in Fig.~\ref{fig_fd}(a). In this context, $i\in\mathcal{N}_{+x}$ out of $N$ pedestrians are walking in the $+x$ direction, and the remaining $i\in\mathcal{N}_{-x}$ pedestrians are traveling in the $-x$ direction. Here, we assume $n(\mathcal{N}_{+x})=n(\mathcal{N}_{-x})=N/2$. We also assume that the periodic boundary condition in the $x$ direction, which is set so that the global average density of $\rho_\mathrm{av}=N/(LW)$, is constant. Such situations, in which a self-organizing lane-formation and clogging phenomenon occurs at high-density points (which is often called jamming-transition), have been extensively studied ~\cite{HFV2000a,Nagatani2009,ZKS+2012,FN2016,FMN2018}. 

According to the experiments examining the impacts of congestion in a corridor, pedestrian flow velocities tend to decrease as density increases (see, e.g., Ref.~\cite{ZKS+2012} and Fig.~S1 in the Supplemental Information (SI)). In the present study, we consider situations in which an obstacle array is placed along the median line of a corridor to suppress velocity reductions and to control pedestrian flow patterns. 
To be more specific, we consider elliptic obstacles with major and minor axes of lengths $2a$ and $2b$, respectively, which are placed on the median line of the corridor ($y=0$) at intervals of $L_p$. The obstacles are commonly angled, such that the angle between the $x$ axis and the major axis is $\varphi$.

%

%
%
\section{\label{sec_md}Molecular dynamics simulation}
%
%

Before showing simulation results, we will first summarize the model equations used in the molecular dynamics. Each pedestrian is modeled by a spherical particle, the dynamics of which is governed by the following equation of motion:
\begin{align}
m_i\frac{\mathrm{d}\bm{v}_i}{\mathrm{d}t} = &-m_i\frac{\bm{v}_i - v_d \bm{e}_i}{\tau} + \sum_{j \neq i} \bm{f}_{ij} 
+ \sum_{k\in \textrm{wall}}\bm{f}^w_{ik} + \bm{\xi}_i,
\label{eq_sfm}
\end{align}
where $m_i$, $r_i$, and $\bm{v}_i$ are the mass, radius, and velocity of the $i$th particle, respectively. The first term on the right-hand side represents the force driving the pedestrian in the desired direction $\bm{e}_i$ with velocity $v_d$ and relaxation time $\tau$. The second term is the sum of the pairwise interaction force $\bm{f}_{ij}$ between pedestrian particles $i$ and $j$. In the third term, the walls and obstacles are expressed in terms of groups of fixed particles, indexed by $k\in$ wall, with $\bm{f}^w_{ik}$ being the interaction force between particle $i$ and those fixed particles (see Fig.~S2 in the SI). Finally, $\bm{\xi}_i$ indicates the 
Gaussian white 
noise satisfying $\langle\bm{\xi}_i\rangle=\bm{0}$, $\langle\xi_i(t)\xi_i(t')\rangle=\Xi\bm{I}\delta(t-t')$, where $\delta$ is the Kronecker delta, $\bm{I}$ is the identity matrix, and $\Xi$ is the noise intensity parameter.

The explicit form of $\bm{f}_{ij}$ is given by
\begin{align}
\bm{f}_{ij} = &[Ae^{-r'_{ij}/B} - \kappa r'_{ij}u(-r'_{ij})]\bm{n}_{ij} \nonumber\\
&\qquad \qquad \qquad - gr'_{ij}u(-r'_{ij})\Delta v_{ij}^t \bm{t}_{ij},
\label{eq_pp}
\end{align}
where $A$, $B$, $\kappa$, and $g$ are the model parameters, and $r'_{ij}=r_{ij}-(r_i + r_j)$, with $r_{ij}$ being the distance between particle $i$ and $j$, and $u(z)$ is the Heaviside function, the value of which is unity for $z>0$ and zero otherwise. The normal unit vector $\bm{n}_{ij}$ is pointing from the position of particle $j$ to that of $i$, and the unit vector $\bm{t}_{ij}$ is in the tangential direction perpendicular to $\bm{n}_{ij}$. $\Delta v_{ij}^t$ is the projection of the relative velocity between $i$ and $j$ on $\bm{t}_{ij}$. The interaction force $\bm{f}^w_{ik}$ has the same form as Eq.~\eqref{eq_pp} with the parameters $A$ and $B$ simply replaced by $A_w$ and $B_w$.
\begin{figure}[t]
\begin{center}
\includegraphics[width=0.9\hsize]{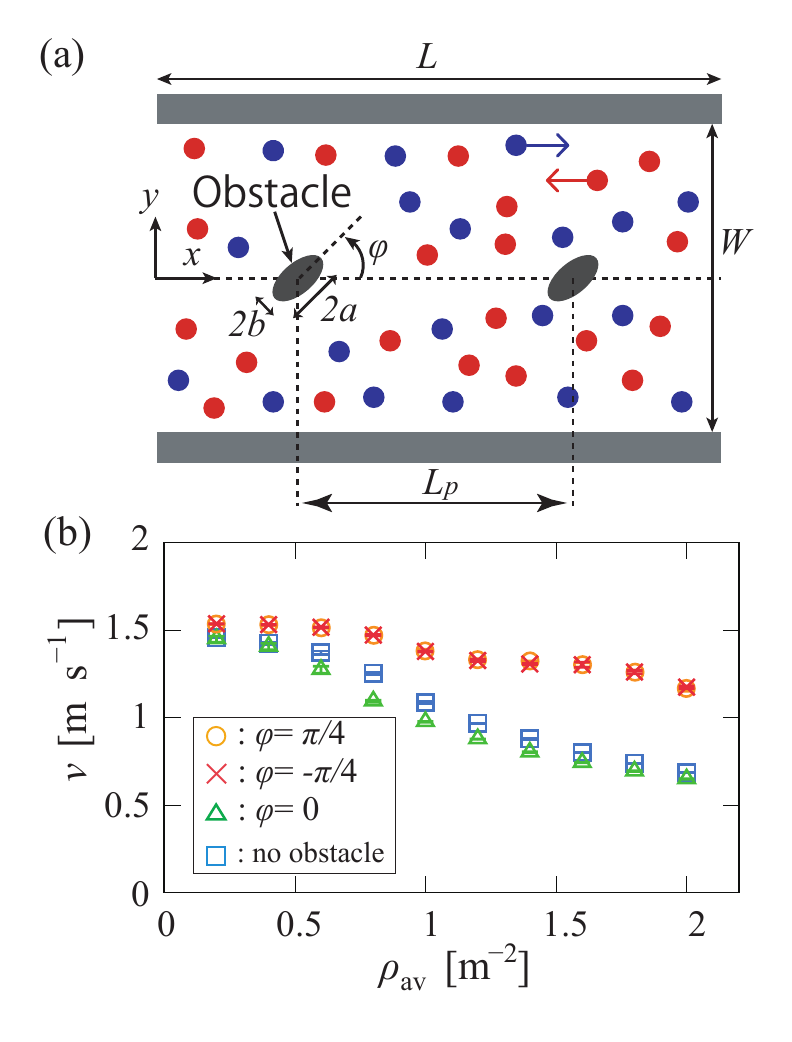}
\caption{The present system and fundamental diagram. (a) Schematic of the geometry. (b) Average velocity $v$ vs average density $\rho_\mathrm{av}$ obtained with the simulations for different situations. The cases with obstacles of $\varphi=0$ and $\pm\pi/4$ are shown along with the no-obstacle case. The symbols with an error bar indicate the mean value with the standard deviations of no fewer than five samples with different initial conditions. For each run, the value of $v$ is the average over $1.5\times 10^7$ steps 
with time step $dt=0.001$\,s, i.e., $1.5 \times 10^4$\,s
.}
\label{fig_fd}
\end{center}
\end{figure}
%

All the simulations are implemented using the open source code LAMMPS~\cite{lammps}.
The source codes in the original package are modified to incorporate 
the pairwise interactions corresponding to Eq.~(2) (see Sec.~S1 of the SI for details). 
The specific values of parameters used in our simulations are summarized as shown below. The mass of each pedestrian is $m_i=80$\,kg and the diameter of a pedestrian $d_p=2r_i$ is $0.3$\,m. We choose $d_w = 1/2\sqrt{2}$\,m as the diameter of fixed particles. 
The model parameters for the pairwise interaction force given in Eq.~(\ref{eq_pp}) are fixed at $A=A_w=2000$\,N, $B=B_w=0.08$\,m, $\kappa=1.2\times 10^5$\,N/m, $g=2.4\times 10^5$\,Pa$\cdot$s, and $\tau=0.5$\,s,
following Ref.~\cite{HFV2000}. The repulsive force is taken into account only for $r_{ij}< 3.0$\,m, and is otherwise cut off. The value of pedestrians' desired velocity or terminal velocity is set as $v_d=1.55$\,m/s, which is based on the experimental result in Ref.~\cite {ZKS+2012}, and the noise intensity is chosen as $\Xi=6.63\times 10^5$\, N$^2$ to reproduce the experimental density-velocity relationship discussed below (see Fig.~S3 in the SI). The initial configuration is constructed with randomly distributed pedestrians, and each simulation runs over $2\times 10^7$ steps with time step $dt = 0.001$\,s. Note that the simulations set with these parameters reproduce the experimental results well, as shown in Fig.~S1 in the SI. In the following simulation results, the geometrical parameters are fixed at $L=20$\,m,\,$W=8$\,m,\,$a=0.7$\,m,\,$b=0.4$\,m, and $L_p=10$\,m unless otherwise stated.
%

We show in Fig.~\ref{fig_fd}(b) the {\it fundamental diagram}, namely the density-velocity relation for our system. More precisely, we plot the velocity averaged over the pedestrians walking in the $+x$ direction versus the average density $\rho_{\mathrm{av}}$ ($=N/LW$). Here, and in what follows, the time average is taken over $1.5\times 10^7$ steps for each run, and no fewer than five runs with different initial configurations are used to obtain each averaged quantity. 
In general, pedestrian counter flows under ordinary situations in the absence of obstacles exhibit congestion as density increases, resulting in monotonically decreasing velocity (see, e.g., Ref.~\cite{ZKS+2012}). This feature is properly captured by our simulation for the no-obstacle case results, which are shown as a reference in Fig.~\ref{fig_fd}(b). In the same figure, the case in the presence of obstacles with $\varphi=0$, i.e., the symmetric obstacles, which have no impact on this fundamental diagram, is also shown. 
On the other hand, the corridor with emplaced asymmetric obstacles ($\varphi=-\pi/4$ and $\pi/4$) maintains a much higher velocity than that in the previous two cases. The simulation snapshots (see Fig.~S4 in the SI) imply that this significant velocity enhancement (thus, flux) is a result of lane formations that reduce friction between particles passing in opposite directions. 
We also note here that the formed lanes are stable. In other words, once a lane is formed, it tends to occupy the same side of the corridor for a long time. In our simulations, the lanes with emplaced asymmetric obstacles do not change sides during the simulation runs (see Fig.~S5 in the SI). 
At this point, we see that the flow structure, i.e., whether or not lanes are formed, plays an essential role in improving traffic flow efficiency.

Next, we investigate the structure of the pedestrian counter flows in the presence of the obstacles in greater detail. In order to quantify the flow structure discussed above, we introduce the order parameter $\Phi$, which is defined as
\begin{align}
\Phi = \frac{1}{N} \sum_{i=1}^{N} \frac{v_{xi} \cdot y_i}{|v_{xi} \cdot y_i|},
\label{eq_phi}
\end{align}
where $v_{xi}$ and $y_i$ are the $x$ component of the velocity and the $y$ component of the position of particle $i$, respectively~\cite{OVH+2016}.
Since $y=0$ is the median line of the corridor, the value of $v_{xi}\cdot y_i$ is positive when particle $i$ moves in the $+x$ direction in the region $y>0$. Therefore, $\Phi>0$ when most pedestrians keep to their left, and similarly $\Phi<0$ if they keep to their right. The value of $\Phi$ vanishes when the pedestrians walking in the opposite directions are uniformly distributed or when the keep-left and keep-right patterns appear with equal probability. This order parameter is normalized such that $|\Phi|=1$ when the pedestrian flow is perfectly separated into two streams (see Fig.~\ref{fig_cf}). 

\begin{figure}[t]
\begin{center}
\includegraphics[width=0.8\hsize]{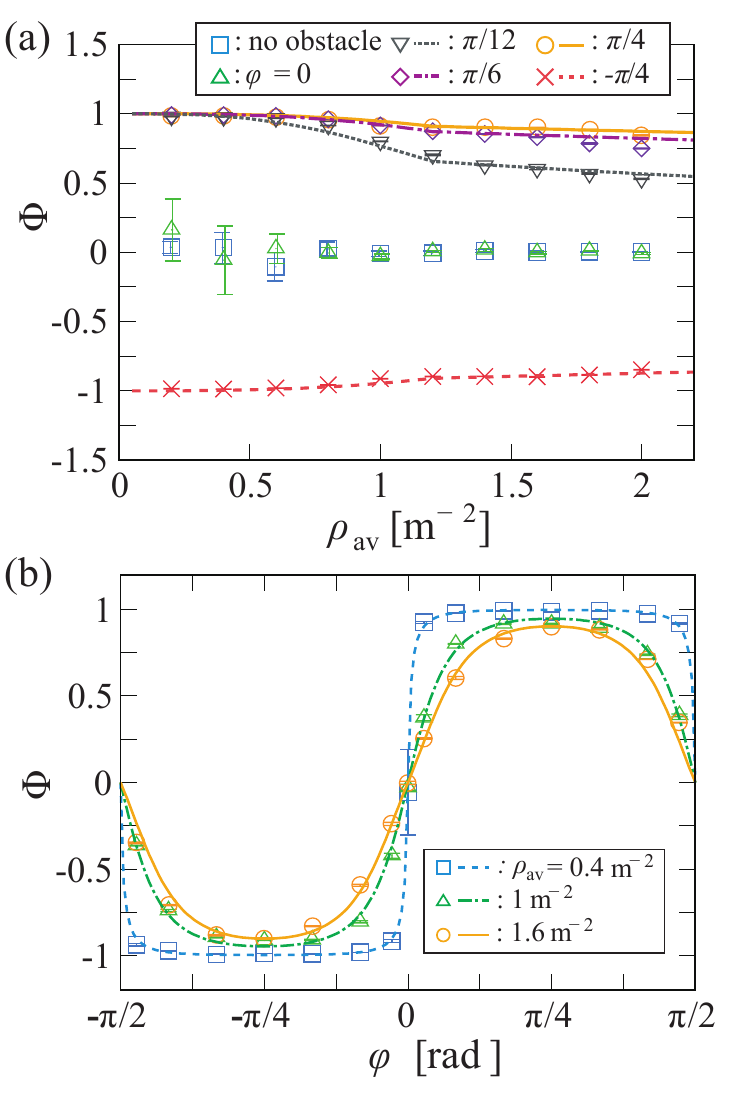}
\caption{Order parameter $\Phi$ (a) as a function of average density $\rho_{\mathrm{av}}$ for various obstacles, and (b) as a function of angle $\varphi$ of obstacles for various values of $\rho_{\mathrm{av}}$. The symbols indicate the simulation results (see the caption of Fig.~\ref{fig_fd}).
The predictions of the model given by Eq.~\eqref{eq_diff} are shown with lines.}
\label{fig_phis}
\end{center}
\end{figure}

In Fig.~\ref{fig_phis}(a), we show the order parameter $\Phi$ as a function of the average density $\rho_{\mathrm{av}}$ for the situations considered in Fig.~\ref{fig_fd}(b), supplemented by the cases of $\varphi=\pi/6$ and \,$\pi/12$. In the absence of obstacles, the system is purely symmetric about $y=0$. Hence, we see $\Phi \sim 0$ in the entire range of $\rho_{\mathrm{av}}$. Again, the symmetric obstacles with $\varphi=0$ do not influence the flow structure. On the other hand, when the obstacles are angled by $\varphi=\pm\pi/4$, $|\Phi|\sim 1$ for the wide range up to $\rho_\mathrm{av}\sim 1$\,m$^{-2}$, and $|\Phi|$ is larger than $0.75$ even for the high densities. In the case of shallower $|\varphi|<\pi/4$ angles, the absolute value of $\Phi$ is smaller, but remains significant. In other words, the pedestrian flow exhibits clear self-organization by separating into two groups that keep to their left or to their right as they travel in opposite directions.

In view of the fact that the separation does not occur in the absence of  obstacles,
the lane formation obtained here is in contrast to that observed in a bulk situation, i.e., in a system without obstacles, for certain parameter ranges (see, e.g., \cite{IK2017,GL2012,IWH2012,RTP+2018}).
In the present case, the asymmetry of obstacles is the main contribution
to the lane formation and stabilization. This is confirmed by the symmetric obstacle results, which still do not lead to the stable lane formation.
The local interaction with a tilted obstacle transfers a part of momentum in the $x$ direction into a momentum in the $y$ direction. Depending on the direction (along $x$) from which a particle collides, the gained momentum in the $y$ direction in fact differs. This local imbalance diffuses to the entire region of the corridor, which results in the complete separation
with $\Phi\sim 1$ observed in Fig.~\ref{fig_phis}(a).

%
%

In order to examine the effect of the geometrical parameter in greater detail, we show the relationship between $\Phi$ and $\varphi$ at several values of $\rho_{\mathrm{av}}$ in Fig.~\ref{fig_phis}(b). The order parameter sign is determined by the angle $\varphi$, such that $\Phi>0$ for $\varphi>0$ and $\Phi<0$ for $\varphi<0$, which confirms that tuning the geometrical parameter enables us not only to induce self-organized lane formations but also to control the flow patterns 
precisely, i.e., keep them left or keep them right. Figure~\ref{fig_phis}(b) also shows that the control sensitivity to $\varphi$ depends on the density. More specifically, $|\Phi|$ is rather sensitive for the high density value of $\rho_{\mathrm{av}}=1.6$\,m$^{-2}$, whereas it is robust for the lower density value of $\rho_{\mathrm{av}}=0.4$\,m$^{-2}$. 


%
%

%
\section{\label{sec_md_model}Membrane model for separation}

We next present a model reproducing the separation behavior of the pedestrian counter flows discussed above. We begin with noting that the role played by the obstacles is reminiscent of the effects of filtering membranes used to separate different fluid components. Therefore, inspired by the modeling of such filtering membranes~\cite{MB2017,MB2019}, we now construct a differential equation describing the dynamics of pedestrian concentrations. For simplicity, we assume that the pedestrian density and velocity values are uniform on each side of the membrane. The density of pedestrians walking in the $+x$ or $-x$ direction in the region $y>0$ is written as $\rho^{\pm x}_{y>0}$. Then, the density in the region $y<0$ is $\rho^{\pm x}_{y<0}=\rho_\mathrm{av}/2-\rho^{\pm x}_{y>0}$ because the density of people going in each direction in the whole area of $-W/2<y<W/2$ is $\rho^{+x}=\rho^{-x}=\rho_\mathrm{av}/2$. Furthermore, assuming the density of all pedestrians is uniformly distributed, we have $\rho^{+x}_{y>0}+\rho^{-x}_{y>0}=\rho_\mathrm{av}/2$. Hence, the system density distribution is fully determined once a governing equation for $\rho^{+x}_{y>0}$ is solved. In the following, $\rho^{+x}_{y>0}$ is simply written as $\rho$, and the order parameter is expressed as $\Phi=2(\rho/\rho_\mathrm{av}-1/2)$.

Using the same idea as the model describing concentration variations between two reservoirs separated by a membrane
~\cite{Zwanzig1990,MB2017}
, we model the behavior of $\rho$ as follows:
\begin{align}
\dfrac{\mathrm{d}\rho}{\mathrm{d} t}=\dfrac{M_1L}{eS}\left(\dfrac{\rho_\mathrm{av}}{2}-\rho\right)
+\dfrac{M_2L}{eS}\left(\rho_\mathrm{av}-\rho\right),
\label{eq_diff}
\end{align}
where, $e$ is the thickness of the membrane, $S=LW/2$ is the area of interest (now the region $y>0$).
In contrast to the models describing ordinary membranes~\cite{Zwanzig1990,MB2017},
the equation above contains two terms driving the density change characterized by 
  $M_1$ and $M_2$;
 the parameter $M_1$ controls the driving force that mixes the pedestrians such that the values of $\rho$ in $y>0$ and $y<0$ approach. 
On the other hand, $M_2$ is the parameter for the driving force that separates the pedestrians moving in the $+x$ and $-x$ directions such that $\rho$ approaches $\rho_\mathrm{av}$.
The solution of Eq.~\eqref{eq_diff} is readily obtained as $\rho=\rho_0+\rho_1\exp(-M_1Lt/eS)+\rho_2\exp(-M_2Lt/eS)$,
where $\rho_1$ and $\rho_2$ are the constants determined from the initial conditions and $\rho_0=\rho_{\mathrm{av}}(M_1+2M_2)/(2M_1+2M_2)$ is the stationary solution.

Diffusion phenomena should dominate the physical mechanisms of mixing. 
Since the analysis of counter flows in the bulk region shows generally increasing diffusion coefficient with increasing density, 
we assume a functional form for $M_1$ which increases with density. 
The diffusion coefficient in the $y$ direction obtained under wall-free bulk conditions, in comparison with our model for $M_1$, is found in Fig.~S6 of the SI,
 where the linear time dependent of the mean square displacement shows the ordinary diffusion process in the $y$ direction.
On the other hand, in modeling the separating force $M_2$, we take into account the fact that a certain separation effect is observed even in the low-density region.
Hence we assume a constant value for $M_2$ with respect to $\rho_{\mathrm{av}}$.
However, since the separation ability should depend on the the geometrical details of the obstacles forming the membrane,
we assume that $M_2$ depends on the angle $\varphi$.
We note that a common function of $\varphi$ is assigned to $M_2$ for all the results shown below 
(see Fig.~S7 in the SI,) of which the functional form reflects the shape of obstacles constituting the membrane.
%
%
The values of $\Phi$ obtained from the steady solution of Eq.~\eqref{eq_diff} are shown in Figs.~\ref{fig_phis}(a) and (b). The results of molecular dynamics simulations, including the variation of decreasing $|\Phi|$ in Fig.~\ref{fig_phis}(a), are well captured. 
In addition, 
the molecular dynamics results examining the effect of the interval between obstacles $L_p$ are well predicted by the present model with the same parameter set (see Fig.~S8 in the SI).

From the above comparison, we conclude that the minimum model given in Eq.~\eqref{eq_diff}, based on the membrane dynamics (with the driving forces chosen appropriately), is capable of predicting the spontaneous separation effect of the pedestrian flow. In addition to the steady-state behavior focused on in the above discussion, the present model was also examined for transient response, and its consistency with the molecular dynamics simulations was confirmed to be within the parameter range where the approximation of Eq.~\eqref{eq_diff} is valid. In other words, situations where the corridor area is not too wide and the density to each side of the membrane varies uniformly. More specifically, for the molecular dynamics simulation, we measure the relaxation time taken to reach the steady-state exhibiting $|\Phi|\sim 1$, starting from the initial condition with $\Phi=0$, in which the pedestrians traveling in both directions distribute uniformly in the whole corridor. These results are then compared with the corresponding relaxation time predicted by the model given in Eq.~\eqref{eq_diff}. The resulting model predictions agree well with the molecular dynamics results under conditions in which $W$ is not too large (see Fig.~S9 in the SI).

%

%
%
\section{\label{sec_conclusion}Conclusion}

To summarize, we have shown that simple asymmetric obstacles emplaced in a corridor enable us to control pedestrian flow patterns by inducing self-organizing lane formations. As demonstrated in Fig.~\ref{fig_fd}, the structured pedestrian flows are more efficient than those of ordinary unstructured crowds. Thus, the present results could contribute to developing new concepts for engineering corridor designs in ways that create efficient traffic lanes. Here, the asymmetry is generated by designing the geometrical shape of obstacles. However, this is merely an example for realizing the spontaneous separation of pedestrians. 
The original concept of the ``social force'' employed in our molecular dynamics simulations includes psychological interactions forces acting effectively on the pedestrians. Therefore, designing {\it psychological obstacles}, constructed by means of visual effects such as photo-regulation, electronic signage, or some other methods, could provide alternative approaches, and will be included among our future research topics.

Our membrane model for separation has shown good agreements with the molecular dynamics simulation results, as shown in Fig.~\ref{fig_phis}. This is an example of the analogies found in multidisciplinary studies, in which a theory established for microscopic physics is used to explain macroscopic phenomena at a length scale that is orders of magnitude higher than the original micro-scale. We believe that our finding suggests a way forward for the field of mobility and transportation, particularly when viewed in tandem with ideas on various unobvious phenomena present in microscopic transportation systems.

%
\section*{Acknowledgments}
The authors would like to thank Dr. K. Kidono of Toyota Central R\&D Labs., Inc. 
and M. Shimada of the University of Tokyo for the useful discussions.


\begin{thebibliography}{47}%
\makeatletter
\providecommand \@ifxundefined [1]{%
 \@ifx{#1\undefined}
}%
\providecommand \@ifnum [1]{%
 \ifnum #1\expandafter \@firstoftwo
 \else \expandafter \@secondoftwo
 \fi
}%
\providecommand \@ifx [1]{%
 \ifx #1\expandafter \@firstoftwo
 \else \expandafter \@secondoftwo
 \fi
}%
\providecommand \natexlab [1]{#1}%
\providecommand \enquote  [1]{``#1''}%
\providecommand \bibnamefont  [1]{#1}%
\providecommand \bibfnamefont [1]{#1}%
\providecommand \citenamefont [1]{#1}%
\providecommand \href@noop [0]{\@secondoftwo}%
\providecommand \href [0]{\begingroup \@sanitize@url \@href}%
\providecommand \@href[1]{\@@startlink{#1}\@@href}%
\providecommand \@@href[1]{\endgroup#1\@@endlink}%
\providecommand \@sanitize@url [0]{\catcode `\\12\catcode `\$12\catcode
  `\&12\catcode `\#12\catcode `\^12\catcode `\_12\catcode `\%12\relax}%
\providecommand \@@startlink[1]{}%
\providecommand \@@endlink[0]{}%
\providecommand \url  [0]{\begingroup\@sanitize@url \@url }%
\providecommand \@url [1]{\endgroup\@href {#1}{\urlprefix }}%
\providecommand \urlprefix  [0]{URL }%
\providecommand \Eprint [0]{\href }%
\providecommand \doibase [0]{http://dx.doi.org/}%
\providecommand \selectlanguage [0]{\@gobble}%
\providecommand \bibinfo  [0]{\@secondoftwo}%
\providecommand \bibfield  [0]{\@secondoftwo}%
\providecommand \translation [1]{[#1]}%
\providecommand \BibitemOpen [0]{}%
\providecommand \bibitemStop [0]{}%
\providecommand \bibitemNoStop [0]{.\EOS\space}%
\providecommand \EOS [0]{\spacefactor3000\relax}%
\providecommand \BibitemShut  [1]{\csname bibitem#1\endcsname}%
\let\auto@bib@innerbib\@empty
\bibitem [{\citenamefont {Barbosa}\ \emph {et~al.}(2018)\citenamefont
  {Barbosa}, \citenamefont {Barthelemy}, \citenamefont {Ghoshal}, \citenamefont
  {James}, \citenamefont {Lenormand}, \citenamefont {Louail}, \citenamefont
  {Menezes}, \citenamefont {Ramasco}, \citenamefont {Simini},\ and\
  \citenamefont {Tomasini}}]{BBG+2018}%
  \BibitemOpen
  \bibfield  {author} {\bibinfo {author} {\bibfnamefont {Hugo}\ \bibnamefont
  {Barbosa}}, \bibinfo {author} {\bibfnamefont {Marc}\ \bibnamefont
  {Barthelemy}}, \bibinfo {author} {\bibfnamefont {Gourab}\ \bibnamefont
  {Ghoshal}}, \bibinfo {author} {\bibfnamefont {Charlotte~R}\ \bibnamefont
  {James}}, \bibinfo {author} {\bibfnamefont {Maxime}\ \bibnamefont
  {Lenormand}}, \bibinfo {author} {\bibfnamefont {Thomas}\ \bibnamefont
  {Louail}}, \bibinfo {author} {\bibfnamefont {Ronaldo}\ \bibnamefont
  {Menezes}}, \bibinfo {author} {\bibfnamefont {Jos{\'e}~J}\ \bibnamefont
  {Ramasco}}, \bibinfo {author} {\bibfnamefont {Filippo}\ \bibnamefont
  {Simini}}, \ and\ \bibinfo {author} {\bibfnamefont {Marcello}\ \bibnamefont
  {Tomasini}},\ }\bibfield  {title} {\enquote {\bibinfo {title} {Human
  mobility: Models and applications},}\ }\href@noop {} {\bibfield  {journal}
  {\bibinfo  {journal} {Phys. Rep.}\ }\textbf {\bibinfo {volume} {734}},\
  \bibinfo {pages} {1--74} (\bibinfo {year} {2018})}\BibitemShut {NoStop}%
\bibitem [{\citenamefont {Barthelemy}(2019)}]{Barthelemy2019}%
  \BibitemOpen
  \bibfield  {author} {\bibinfo {author} {\bibfnamefont {Marc}\ \bibnamefont
  {Barthelemy}},\ }\bibfield  {title} {\enquote {\bibinfo {title} {The
  statistical physics of cities},}\ }\href@noop {} {\bibfield  {journal}
  {\bibinfo  {journal} {Nature Rev. Phys.}\ ,\ \bibinfo {pages} {1}} (\bibinfo
  {year} {2019})}\BibitemShut {NoStop}%
\bibitem [{\citenamefont {Lv}\ \emph {et~al.}(2014)\citenamefont {Lv},
  \citenamefont {Duan}, \citenamefont {Kang}, \citenamefont {Li},\ and\
  \citenamefont {Wang}}]{LDK+2014}%
  \BibitemOpen
  \bibfield  {author} {\bibinfo {author} {\bibfnamefont {Yisheng}\ \bibnamefont
  {Lv}}, \bibinfo {author} {\bibfnamefont {Yanjie}\ \bibnamefont {Duan}},
  \bibinfo {author} {\bibfnamefont {Wenwen}\ \bibnamefont {Kang}}, \bibinfo
  {author} {\bibfnamefont {Zhengxi}\ \bibnamefont {Li}}, \ and\ \bibinfo
  {author} {\bibfnamefont {Fei-Yue}\ \bibnamefont {Wang}},\ }\bibfield  {title}
  {\enquote {\bibinfo {title} {Traffic flow prediction with big data: a deep
  learning approach},}\ }\href@noop {} {\bibfield  {journal} {\bibinfo
  {journal} {IEEE T. Intell. Transp.}\ }\textbf {\bibinfo {volume} {16}},\
  \bibinfo {pages} {865--873} (\bibinfo {year} {2014})}\BibitemShut {NoStop}%
\bibitem [{\citenamefont {Varga}\ \emph {et~al.}(2016)\citenamefont {Varga},
  \citenamefont {Kov\'acs}, \citenamefont {T\'oth}, \citenamefont {Papp},\ and\
  \citenamefont {N\'eda}}]{VKT+2016}%
  \BibitemOpen
  \bibfield  {author} {\bibinfo {author} {\bibfnamefont {Levente}\ \bibnamefont
  {Varga}}, \bibinfo {author} {\bibfnamefont {Andr\'as}\ \bibnamefont
  {Kov\'acs}}, \bibinfo {author} {\bibfnamefont {Geza}\ \bibnamefont {T\'oth}},
  \bibinfo {author} {\bibfnamefont {Istv\'an}\ \bibnamefont {Papp}}, \ and\
  \bibinfo {author} {\bibfnamefont {Zolt\'an}\ \bibnamefont {N\'eda}},\
  }\bibfield  {title} {\enquote {\bibinfo {title} {Further we travel the faster
  we go},}\ }\href@noop {} {\bibfield  {journal} {\bibinfo  {journal} {PloS
  One}\ }\textbf {\bibinfo {volume} {11}},\ \bibinfo {pages} {e0148913}
  (\bibinfo {year} {2016})}\BibitemShut {NoStop}%
\bibitem [{\citenamefont {Scott}(1974)}]{Scott1974}%
  \BibitemOpen
  \bibfield  {author} {\bibinfo {author} {\bibfnamefont {Allen~J}\ \bibnamefont
  {Scott}},\ }\bibfield  {title} {\enquote {\bibinfo {title} {A theoretical
  model of pedestrian flow},}\ }\href@noop {} {\bibfield  {journal} {\bibinfo
  {journal} {Socio-Econ. Plan. Sci.}\ }\textbf {\bibinfo {volume} {8}},\
  \bibinfo {pages} {317--322} (\bibinfo {year} {1974})}\BibitemShut {NoStop}%
\bibitem [{\citenamefont {Polus}\ \emph {et~al.}(1983)\citenamefont {Polus},
  \citenamefont {Schofer},\ and\ \citenamefont {Ushpiz}}]{PSU1983}%
  \BibitemOpen
  \bibfield  {author} {\bibinfo {author} {\bibfnamefont {Abishai}\ \bibnamefont
  {Polus}}, \bibinfo {author} {\bibfnamefont {Joseph~L}\ \bibnamefont
  {Schofer}}, \ and\ \bibinfo {author} {\bibfnamefont {Ariela}\ \bibnamefont
  {Ushpiz}},\ }\bibfield  {title} {\enquote {\bibinfo {title} {Pedestrian flow
  and level of service},}\ }\href@noop {} {\bibfield  {journal} {\bibinfo
  {journal} {J. Transp. Eng.}\ }\textbf {\bibinfo {volume} {109}},\ \bibinfo
  {pages} {46--56} (\bibinfo {year} {1983})}\BibitemShut {NoStop}%
\bibitem [{\citenamefont {Gipps}\ and\ \citenamefont
  {Marksj{\"o}}(1985)}]{GM1985}%
  \BibitemOpen
  \bibfield  {author} {\bibinfo {author} {\bibfnamefont {Peter~G}\ \bibnamefont
  {Gipps}}\ and\ \bibinfo {author} {\bibfnamefont {B}~\bibnamefont
  {Marksj{\"o}}},\ }\bibfield  {title} {\enquote {\bibinfo {title} {A
  micro-simulation model for pedestrian flows},}\ }\href@noop {} {\bibfield
  {journal} {\bibinfo  {journal} {Math. Comput. Simulat.}\ }\textbf {\bibinfo
  {volume} {27}},\ \bibinfo {pages} {95--105} (\bibinfo {year}
  {1985})}\BibitemShut {NoStop}%
\bibitem [{\citenamefont {Virkler}\ and\ \citenamefont
  {Elayadath}(1994)}]{VE1994}%
  \BibitemOpen
  \bibfield  {author} {\bibinfo {author} {\bibfnamefont {Mark~R}\ \bibnamefont
  {Virkler}}\ and\ \bibinfo {author} {\bibfnamefont {Sathish}\ \bibnamefont
  {Elayadath}},\ }\bibfield  {title} {\enquote {\bibinfo {title} {Pedestrian
  speed-flow-density relationships},}\ }\href@noop {} {\bibfield  {journal}
  {\bibinfo  {journal} {Transp. Res. Rec.}\ }\textbf {\bibinfo {volume}
  {1438}},\ \bibinfo {pages} {51--18} (\bibinfo {year} {1994})}\BibitemShut
  {NoStop}%
\bibitem [{\citenamefont {Seyfried}\ \emph {et~al.}(2005)\citenamefont
  {Seyfried}, \citenamefont {Steffen}, \citenamefont {Klingsch},\ and\
  \citenamefont {Boltes}}]{SSK+2005}%
  \BibitemOpen
  \bibfield  {author} {\bibinfo {author} {\bibfnamefont {Armin}\ \bibnamefont
  {Seyfried}}, \bibinfo {author} {\bibfnamefont {Bernhard}\ \bibnamefont
  {Steffen}}, \bibinfo {author} {\bibfnamefont {Wolfram}\ \bibnamefont
  {Klingsch}}, \ and\ \bibinfo {author} {\bibfnamefont {Maik}\ \bibnamefont
  {Boltes}},\ }\bibfield  {title} {\enquote {\bibinfo {title} {The fundamental
  diagram of pedestrian movement revisited},}\ }\href@noop {} {\bibfield
  {journal} {\bibinfo  {journal} {J. Stat. Mech.}\ }\textbf {\bibinfo {volume}
  {2005}},\ \bibinfo {pages} {P10002} (\bibinfo {year} {2005})}\BibitemShut
  {NoStop}%
\bibitem [{\citenamefont {Helbing}\ \emph {et~al.}(2007)\citenamefont
  {Helbing}, \citenamefont {Johansson},\ and\ \citenamefont
  {Al-Abideen}}]{HJA2007}%
  \BibitemOpen
  \bibfield  {author} {\bibinfo {author} {\bibfnamefont {Dirk}\ \bibnamefont
  {Helbing}}, \bibinfo {author} {\bibfnamefont {Anders}\ \bibnamefont
  {Johansson}}, \ and\ \bibinfo {author} {\bibfnamefont {Habib~Zein}\
  \bibnamefont {Al-Abideen}},\ }\bibfield  {title} {\enquote {\bibinfo {title}
  {Dynamics of crowd disasters: An empirical study},}\ }\href@noop {}
  {\bibfield  {journal} {\bibinfo  {journal} {Phys. Rev. E}\ }\textbf {\bibinfo
  {volume} {75}},\ \bibinfo {pages} {046109} (\bibinfo {year}
  {2007})}\BibitemShut {NoStop}%
\bibitem [{\citenamefont {Johansson}\ \emph {et~al.}(2008)\citenamefont
  {Johansson}, \citenamefont {Helbing}, \citenamefont {Al-Abideen},\ and\
  \citenamefont {Al-Bosta}}]{JHA+2008}%
  \BibitemOpen
  \bibfield  {author} {\bibinfo {author} {\bibfnamefont {Anders}\ \bibnamefont
  {Johansson}}, \bibinfo {author} {\bibfnamefont {Dirk}\ \bibnamefont
  {Helbing}}, \bibinfo {author} {\bibfnamefont {Habib~Z}\ \bibnamefont
  {Al-Abideen}}, \ and\ \bibinfo {author} {\bibfnamefont {Salim}\ \bibnamefont
  {Al-Bosta}},\ }\bibfield  {title} {\enquote {\bibinfo {title} {From crowd
  dynamics to crowd safety: a video-based analysis},}\ }\href@noop {}
  {\bibfield  {journal} {\bibinfo  {journal} {Adv. Complex Sys.}\ }\textbf
  {\bibinfo {volume} {11}},\ \bibinfo {pages} {497--527} (\bibinfo {year}
  {2008})}\BibitemShut {NoStop}%
\bibitem [{\citenamefont {Chattaraj}\ \emph {et~al.}(2009)\citenamefont
  {Chattaraj}, \citenamefont {Seyfried},\ and\ \citenamefont
  {Chakroborty}}]{CSC2009}%
  \BibitemOpen
  \bibfield  {author} {\bibinfo {author} {\bibfnamefont {Ujjal}\ \bibnamefont
  {Chattaraj}}, \bibinfo {author} {\bibfnamefont {Armin}\ \bibnamefont
  {Seyfried}}, \ and\ \bibinfo {author} {\bibfnamefont {Partha}\ \bibnamefont
  {Chakroborty}},\ }\bibfield  {title} {\enquote {\bibinfo {title} {Comparison
  of pedestrian fundamental diagram across cultures},}\ }\href@noop {}
  {\bibfield  {journal} {\bibinfo  {journal} {Adv. Complex Sys.}\ }\textbf
  {\bibinfo {volume} {12}},\ \bibinfo {pages} {393--405} (\bibinfo {year}
  {2009})}\BibitemShut {NoStop}%
\bibitem [{\citenamefont {Moussa{\"\i}d}\ \emph {et~al.}(2009)\citenamefont
  {Moussa{\"\i}d}, \citenamefont {Helbing}, \citenamefont {Garnier},
  \citenamefont {Johansson}, \citenamefont {Combe},\ and\ \citenamefont
  {Theraulaz}}]{MHG+2009}%
  \BibitemOpen
  \bibfield  {author} {\bibinfo {author} {\bibfnamefont {Mehdi}\ \bibnamefont
  {Moussa{\"\i}d}}, \bibinfo {author} {\bibfnamefont {Dirk}\ \bibnamefont
  {Helbing}}, \bibinfo {author} {\bibfnamefont {Simon}\ \bibnamefont
  {Garnier}}, \bibinfo {author} {\bibfnamefont {Anders}\ \bibnamefont
  {Johansson}}, \bibinfo {author} {\bibfnamefont {Maud}\ \bibnamefont {Combe}},
  \ and\ \bibinfo {author} {\bibfnamefont {Guy}\ \bibnamefont {Theraulaz}},\
  }\bibfield  {title} {\enquote {\bibinfo {title} {Experimental study of the
  behavioural mechanisms underlying self-organization in human crowds},}\
  }\href@noop {} {\bibfield  {journal} {\bibinfo  {journal} {Proc. R. Soc.
  Lond. B}\ }\textbf {\bibinfo {volume} {276}},\ \bibinfo {pages} {2755--2762}
  (\bibinfo {year} {2009})}\BibitemShut {NoStop}%
\bibitem [{\citenamefont {Zanlungo}\ \emph {et~al.}(2011)\citenamefont
  {Zanlungo}, \citenamefont {Ikeda},\ and\ \citenamefont {Kanda}}]{ZIK2011}%
  \BibitemOpen
  \bibfield  {author} {\bibinfo {author} {\bibfnamefont {Francesco}\
  \bibnamefont {Zanlungo}}, \bibinfo {author} {\bibfnamefont {Tetsushi}\
  \bibnamefont {Ikeda}}, \ and\ \bibinfo {author} {\bibfnamefont {Takayuki}\
  \bibnamefont {Kanda}},\ }\bibfield  {title} {\enquote {\bibinfo {title}
  {Social force model with explicit collision prediction},}\ }\href@noop {}
  {\bibfield  {journal} {\bibinfo  {journal} {Europhys. Lett.}\ }\textbf
  {\bibinfo {volume} {93}},\ \bibinfo {pages} {68005} (\bibinfo {year}
  {2011})}\BibitemShut {NoStop}%
\bibitem [{\citenamefont {Zhang}\ \emph {et~al.}(2011)\citenamefont {Zhang},
  \citenamefont {Klingsch}, \citenamefont {Schadschneider},\ and\ \citenamefont
  {Seyfried}}]{ZKS+2011}%
  \BibitemOpen
  \bibfield  {author} {\bibinfo {author} {\bibfnamefont {Jun}\ \bibnamefont
  {Zhang}}, \bibinfo {author} {\bibfnamefont {Wolfram}\ \bibnamefont
  {Klingsch}}, \bibinfo {author} {\bibfnamefont {Andreas}\ \bibnamefont
  {Schadschneider}}, \ and\ \bibinfo {author} {\bibfnamefont {Armin}\
  \bibnamefont {Seyfried}},\ }\bibfield  {title} {\enquote {\bibinfo {title}
  {Transitions in pedestrian fundamental diagrams of straight corridors and
  {T}-junctions},}\ }\href@noop {} {\bibfield  {journal} {\bibinfo  {journal}
  {J. Stat. Mech. Theory E.}\ }\textbf {\bibinfo {volume} {2011}},\ \bibinfo
  {pages} {P06004} (\bibinfo {year} {2011})}\BibitemShut {NoStop}%
\bibitem [{\citenamefont {Zhang}\ \emph {et~al.}(2012)\citenamefont {Zhang},
  \citenamefont {Klingsch}, \citenamefont {Schadschneider},\ and\ \citenamefont
  {Seyfried}}]{ZKS+2012}%
  \BibitemOpen
  \bibfield  {author} {\bibinfo {author} {\bibfnamefont {Jun}\ \bibnamefont
  {Zhang}}, \bibinfo {author} {\bibfnamefont {Wolfram}\ \bibnamefont
  {Klingsch}}, \bibinfo {author} {\bibfnamefont {Andreas}\ \bibnamefont
  {Schadschneider}}, \ and\ \bibinfo {author} {\bibfnamefont {Armin}\
  \bibnamefont {Seyfried}},\ }\bibfield  {title} {\enquote {\bibinfo {title}
  {Ordering in bidirectional pedestrian flows and its influence on the
  fundamental diagram},}\ }\href@noop {} {\bibfield  {journal} {\bibinfo
  {journal} {J. Stat. Mech. Theory E.}\ }\textbf {\bibinfo {volume} {2012}},\
  \bibinfo {pages} {P02002} (\bibinfo {year} {2012})}\BibitemShut {NoStop}%
\bibitem [{\citenamefont {Helbing}(1992)}]{Helbing1992}%
  \BibitemOpen
  \bibfield  {author} {\bibinfo {author} {\bibfnamefont {Dirk}\ \bibnamefont
  {Helbing}},\ }\bibfield  {title} {\enquote {\bibinfo {title} {A fluid dynamic
  model for the movement of pedestrians},}\ }\href@noop {} {\bibfield
  {journal} {\bibinfo  {journal} {Complex Systems}\ }\textbf {\bibinfo {volume}
  {6}},\ \bibinfo {pages} {391--415} (\bibinfo {year} {1992})}\BibitemShut
  {NoStop}%
\bibitem [{\citenamefont {Burstedde}\ \emph {et~al.}(2001)\citenamefont
  {Burstedde}, \citenamefont {Klauck}, \citenamefont {Schadschneider},\ and\
  \citenamefont {Zittartz}}]{BKS+2001}%
  \BibitemOpen
  \bibfield  {author} {\bibinfo {author} {\bibfnamefont {Carsten}\ \bibnamefont
  {Burstedde}}, \bibinfo {author} {\bibfnamefont {Kai}\ \bibnamefont {Klauck}},
  \bibinfo {author} {\bibfnamefont {Andreas}\ \bibnamefont {Schadschneider}}, \
  and\ \bibinfo {author} {\bibfnamefont {Johannes}\ \bibnamefont {Zittartz}},\
  }\bibfield  {title} {\enquote {\bibinfo {title} {Simulation of pedestrian
  dynamics using a two-dimensional cellular automaton},}\ }\href@noop {}
  {\bibfield  {journal} {\bibinfo  {journal} {Physica A}\ }\textbf {\bibinfo
  {volume} {295}},\ \bibinfo {pages} {507--525} (\bibinfo {year}
  {2001})}\BibitemShut {NoStop}%
\bibitem [{\citenamefont {Bonabeau}(2002)}]{Bonabeau2002}%
  \BibitemOpen
  \bibfield  {author} {\bibinfo {author} {\bibfnamefont {Eric}\ \bibnamefont
  {Bonabeau}},\ }\bibfield  {title} {\enquote {\bibinfo {title} {Agent-based
  modeling: Methods and techniques for simulating human systems},}\ }\href@noop
  {} {\bibfield  {journal} {\bibinfo  {journal} {Proc. Natl. Acad. Sci.}\
  }\textbf {\bibinfo {volume} {99}},\ \bibinfo {pages} {7280--7287} (\bibinfo
  {year} {2002})}\BibitemShut {NoStop}%
\bibitem [{\citenamefont {Hughes}(2002)}]{Hughes2002}%
  \BibitemOpen
  \bibfield  {author} {\bibinfo {author} {\bibfnamefont {Roger~L}\ \bibnamefont
  {Hughes}},\ }\bibfield  {title} {\enquote {\bibinfo {title} {A continuum
  theory for the flow of pedestrians},}\ }\href@noop {} {\bibfield  {journal}
  {\bibinfo  {journal} {Transport Res. B-Meth}\ }\textbf {\bibinfo {volume}
  {36}},\ \bibinfo {pages} {507--535} (\bibinfo {year} {2002})}\BibitemShut
  {NoStop}%
\bibitem [{\citenamefont {Seyfried}\ \emph {et~al.}(2006)\citenamefont
  {Seyfried}, \citenamefont {Steffen},\ and\ \citenamefont
  {Lippert}}]{SSL2006}%
  \BibitemOpen
  \bibfield  {author} {\bibinfo {author} {\bibfnamefont {Armin}\ \bibnamefont
  {Seyfried}}, \bibinfo {author} {\bibfnamefont {Bernhard}\ \bibnamefont
  {Steffen}}, \ and\ \bibinfo {author} {\bibfnamefont {Thomas}\ \bibnamefont
  {Lippert}},\ }\bibfield  {title} {\enquote {\bibinfo {title} {Basics of
  modelling the pedestrian flow},}\ }\href@noop {} {\bibfield  {journal}
  {\bibinfo  {journal} {Physica A}\ }\textbf {\bibinfo {volume} {368}},\
  \bibinfo {pages} {232--238} (\bibinfo {year} {2006})}\BibitemShut {NoStop}%
\bibitem [{\citenamefont {Helbing}\ and\ \citenamefont
  {Moln\'ar}(1995)}]{HM1995}%
  \BibitemOpen
  \bibfield  {author} {\bibinfo {author} {\bibfnamefont {Dirk}\ \bibnamefont
  {Helbing}}\ and\ \bibinfo {author} {\bibfnamefont {Peter}\ \bibnamefont
  {Moln\'ar}},\ }\bibfield  {title} {\enquote {\bibinfo {title} {Social force
  model for pedestrian dynamics},}\ }\href@noop {} {\bibfield  {journal}
  {\bibinfo  {journal} {Phys. Rev. E}\ }\textbf {\bibinfo {volume} {51}},\
  \bibinfo {pages} {4282} (\bibinfo {year} {1995})}\BibitemShut {NoStop}%
\bibitem [{\citenamefont {Helbing}\ \emph
  {et~al.}(2000{\natexlab{a}})\citenamefont {Helbing}, \citenamefont {Farkas},\
  and\ \citenamefont {Vicsek}}]{HFV2000}%
  \BibitemOpen
  \bibfield  {author} {\bibinfo {author} {\bibfnamefont {Dirk}\ \bibnamefont
  {Helbing}}, \bibinfo {author} {\bibfnamefont {Ill{\'e}s}\ \bibnamefont
  {Farkas}}, \ and\ \bibinfo {author} {\bibfnamefont {Tamas}\ \bibnamefont
  {Vicsek}},\ }\bibfield  {title} {\enquote {\bibinfo {title} {Simulating
  dynamical features of escape panic},}\ }\href@noop {} {\bibfield  {journal}
  {\bibinfo  {journal} {Nature}\ }\textbf {\bibinfo {volume} {407}},\ \bibinfo
  {pages} {487} (\bibinfo {year} {2000}{\natexlab{a}})}\BibitemShut {NoStop}%
\bibitem [{\citenamefont {Lakoba}\ \emph {et~al.}(2005)\citenamefont {Lakoba},
  \citenamefont {Kaup},\ and\ \citenamefont {Finkelstein}}]{LKF2005}%
  \BibitemOpen
  \bibfield  {author} {\bibinfo {author} {\bibfnamefont {Taras~I}\ \bibnamefont
  {Lakoba}}, \bibinfo {author} {\bibfnamefont {David~J}\ \bibnamefont {Kaup}},
  \ and\ \bibinfo {author} {\bibfnamefont {Neal~M}\ \bibnamefont
  {Finkelstein}},\ }\bibfield  {title} {\enquote {\bibinfo {title}
  {Modifications of the helbing-molnar-farkas-vicsek social force model for
  pedestrian evolution},}\ }\href@noop {} {\bibfield  {journal} {\bibinfo
  {journal} {Simulation}\ }\textbf {\bibinfo {volume} {81}},\ \bibinfo {pages}
  {339--352} (\bibinfo {year} {2005})}\BibitemShut {NoStop}%
\bibitem [{\citenamefont {Oliveira}\ \emph {et~al.}(2016)\citenamefont
  {Oliveira}, \citenamefont {Vieira}, \citenamefont {Helbing}, \citenamefont
  {Andrade~Jr},\ and\ \citenamefont {Herrmann}}]{OVH+2016}%
  \BibitemOpen
  \bibfield  {author} {\bibinfo {author} {\bibfnamefont {C.~L.~N.}\
  \bibnamefont {Oliveira}}, \bibinfo {author} {\bibfnamefont {A.~P.}\
  \bibnamefont {Vieira}}, \bibinfo {author} {\bibfnamefont {Dirk}\ \bibnamefont
  {Helbing}}, \bibinfo {author} {\bibfnamefont {J.~S.}\ \bibnamefont
  {Andrade~Jr}}, \ and\ \bibinfo {author} {\bibfnamefont {Hans~J.}\
  \bibnamefont {Herrmann}},\ }\bibfield  {title} {\enquote {\bibinfo {title}
  {Keep-left behavior induced by asymmetrically profiled walls},}\ }\href@noop
  {} {\bibfield  {journal} {\bibinfo  {journal} {Phys. Rev. X}\ }\textbf
  {\bibinfo {volume} {6}},\ \bibinfo {pages} {011003} (\bibinfo {year}
  {2016})}\BibitemShut {NoStop}%
\bibitem [{\citenamefont {Sticco}\ \emph {et~al.}(2017)\citenamefont {Sticco},
  \citenamefont {Cornes}, \citenamefont {Frank},\ and\ \citenamefont
  {Dorso}}]{SCF+2017}%
  \BibitemOpen
  \bibfield  {author} {\bibinfo {author} {\bibfnamefont {Ignacio~Mariano}\
  \bibnamefont {Sticco}}, \bibinfo {author} {\bibfnamefont {Fernando~Ezequiel}\
  \bibnamefont {Cornes}}, \bibinfo {author} {\bibfnamefont {Guillermo~Alberto}\
  \bibnamefont {Frank}}, \ and\ \bibinfo {author} {\bibfnamefont
  {Claudio~Oscar}\ \bibnamefont {Dorso}},\ }\bibfield  {title} {\enquote
  {\bibinfo {title} {Beyond the faster-is-slower effect},}\ }\href@noop {}
  {\bibfield  {journal} {\bibinfo  {journal} {Phys. Rev. E}\ }\textbf {\bibinfo
  {volume} {96}},\ \bibinfo {pages} {052303} (\bibinfo {year}
  {2017})}\BibitemShut {NoStop}%
\bibitem [{\citenamefont {Kang}\ \emph {et~al.}(2019)\citenamefont {Kang},
  \citenamefont {Zhang},\ and\ \citenamefont {Li}}]{KZL2019}%
  \BibitemOpen
  \bibfield  {author} {\bibinfo {author} {\bibfnamefont {Zengxin}\ \bibnamefont
  {Kang}}, \bibinfo {author} {\bibfnamefont {Lei}\ \bibnamefont {Zhang}}, \
  and\ \bibinfo {author} {\bibfnamefont {Kun}\ \bibnamefont {Li}},\ }\bibfield
  {title} {\enquote {\bibinfo {title} {An improved social force model for
  pedestrian dynamics in shipwrecks},}\ }\href@noop {} {\bibfield  {journal}
  {\bibinfo  {journal} {Appl. Math. Comput.}\ }\textbf {\bibinfo {volume}
  {348}},\ \bibinfo {pages} {355--362} (\bibinfo {year} {2019})}\BibitemShut
  {NoStop}%
\bibitem [{\citenamefont {Ikeda}\ and\ \citenamefont {Kim}(2017)}]{IK2017}%
  \BibitemOpen
  \bibfield  {author} {\bibinfo {author} {\bibfnamefont {Kosuke}\ \bibnamefont
  {Ikeda}}\ and\ \bibinfo {author} {\bibfnamefont {Kang}\ \bibnamefont {Kim}},\
  }\bibfield  {title} {\enquote {\bibinfo {title} {Lane formation dynamics of
  oppositely self-driven binary particles: Effects of density and finite system
  size},}\ }\href@noop {} {\bibfield  {journal} {\bibinfo  {journal} {J. Phys.
  Soc. Jpn.}\ }\textbf {\bibinfo {volume} {86}},\ \bibinfo {pages} {044004}
  (\bibinfo {year} {2017})}\BibitemShut {NoStop}%
\bibitem [{\citenamefont {Feliciani}\ \emph {et~al.}(2018)\citenamefont
  {Feliciani}, \citenamefont {Murakami},\ and\ \citenamefont
  {Nishinari}}]{FMN2018}%
  \BibitemOpen
  \bibfield  {author} {\bibinfo {author} {\bibfnamefont {Claudio}\ \bibnamefont
  {Feliciani}}, \bibinfo {author} {\bibfnamefont {Hisashi}\ \bibnamefont
  {Murakami}}, \ and\ \bibinfo {author} {\bibfnamefont {Katsuhiro}\
  \bibnamefont {Nishinari}},\ }\bibfield  {title} {\enquote {\bibinfo {title}
  {A universal function for capacity of bidirectional pedestrian streams:
  Filling the gaps in the literature},}\ }\href@noop {} {\bibfield  {journal}
  {\bibinfo  {journal} {PLoS One}\ }\textbf {\bibinfo {volume} {13}},\ \bibinfo
  {pages} {e0208496} (\bibinfo {year} {2018})}\BibitemShut {NoStop}%
\bibitem [{\citenamefont {Vissers}\ \emph
  {et~al.}(2011{\natexlab{a}})\citenamefont {Vissers}, \citenamefont {van
  Blaaderen},\ and\ \citenamefont {Imhof}}]{VBI2011}%
  \BibitemOpen
  \bibfield  {author} {\bibinfo {author} {\bibfnamefont {Teun}\ \bibnamefont
  {Vissers}}, \bibinfo {author} {\bibfnamefont {Alfons}\ \bibnamefont {van
  Blaaderen}}, \ and\ \bibinfo {author} {\bibfnamefont {Arnout}\ \bibnamefont
  {Imhof}},\ }\bibfield  {title} {\enquote {\bibinfo {title} {Band formation in
  mixtures of oppositely charged colloids driven by an ac electric field},}\
  }\href@noop {} {\bibfield  {journal} {\bibinfo  {journal} {Phys. Rev. Lett.}\
  }\textbf {\bibinfo {volume} {106}},\ \bibinfo {pages} {228303} (\bibinfo
  {year} {2011}{\natexlab{a}})}\BibitemShut {NoStop}%
\bibitem [{\citenamefont {Vissers}\ \emph
  {et~al.}(2011{\natexlab{b}})\citenamefont {Vissers}, \citenamefont {Wysocki},
  \citenamefont {Rex}, \citenamefont {L{\"o}wen}, \citenamefont {Royall},
  \citenamefont {Imhof},\ and\ \citenamefont {van Blaaderen}}]{VWR+2011}%
  \BibitemOpen
  \bibfield  {author} {\bibinfo {author} {\bibfnamefont {Teun}\ \bibnamefont
  {Vissers}}, \bibinfo {author} {\bibfnamefont {Adam}\ \bibnamefont {Wysocki}},
  \bibinfo {author} {\bibfnamefont {Martin}\ \bibnamefont {Rex}}, \bibinfo
  {author} {\bibfnamefont {Hartmut}\ \bibnamefont {L{\"o}wen}}, \bibinfo
  {author} {\bibfnamefont {C~Patrick}\ \bibnamefont {Royall}}, \bibinfo
  {author} {\bibfnamefont {Arnout}\ \bibnamefont {Imhof}}, \ and\ \bibinfo
  {author} {\bibfnamefont {Alfons}\ \bibnamefont {van Blaaderen}},\ }\bibfield
  {title} {\enquote {\bibinfo {title} {Lane formation in driven mixtures of
  oppositely charged colloids},}\ }\href@noop {} {\bibfield  {journal}
  {\bibinfo  {journal} {Soft Matter}\ }\textbf {\bibinfo {volume} {7}},\
  \bibinfo {pages} {2352--2356} (\bibinfo {year}
  {2011}{\natexlab{b}})}\BibitemShut {NoStop}%
\bibitem [{\citenamefont {Tarama}\ \emph {et~al.}(2019)\citenamefont {Tarama},
  \citenamefont {Egelhaaf},\ and\ \citenamefont {L{\"o}wen}}]{TEL2019}%
  \BibitemOpen
  \bibfield  {author} {\bibinfo {author} {\bibfnamefont {Sonja}\ \bibnamefont
  {Tarama}}, \bibinfo {author} {\bibfnamefont {Stefan~U}\ \bibnamefont
  {Egelhaaf}}, \ and\ \bibinfo {author} {\bibfnamefont {Hartmut}\ \bibnamefont
  {L{\"o}wen}},\ }\bibfield  {title} {\enquote {\bibinfo {title} {Traveling
  band formation in feedback-driven colloids},}\ }\href@noop {} {\bibfield
  {journal} {\bibinfo  {journal} {Phys. Rev. E}\ }\textbf {\bibinfo {volume}
  {100}},\ \bibinfo {pages} {022609} (\bibinfo {year} {2019})}\BibitemShut
  {NoStop}%
\bibitem [{\citenamefont {Kogler}\ and\ \citenamefont {Klapp}(2015)}]{KK2015}%
  \BibitemOpen
  \bibfield  {author} {\bibinfo {author} {\bibfnamefont {Florian}\ \bibnamefont
  {Kogler}}\ and\ \bibinfo {author} {\bibfnamefont {Sabine H.~L.}\ \bibnamefont
  {Klapp}},\ }\bibfield  {title} {\enquote {\bibinfo {title} {Lane formation in
  a system of dipolar microswimmers},}\ }\href@noop {} {\bibfield  {journal}
  {\bibinfo  {journal} {Europhys. Lett.}\ }\textbf {\bibinfo {volume} {110}},\
  \bibinfo {pages} {10004} (\bibinfo {year} {2015})}\BibitemShut {NoStop}%
\bibitem [{\citenamefont {S{\"u}tterlin}\ \emph {et~al.}(2009)\citenamefont
  {S{\"u}tterlin}, \citenamefont {Wysocki}, \citenamefont {Ivlev},
  \citenamefont {R{\"a}th}, \citenamefont {Thomas}, \citenamefont
  {Rubin-Zuzic}, \citenamefont {Goedheer}, \citenamefont {Fortov},
  \citenamefont {Lipaev}, \citenamefont {Molotkov}, \citenamefont {Petrov},
  \citenamefont {Morfill},\ and\ \citenamefont {L\"owen}}]{SWI+2009}%
  \BibitemOpen
  \bibfield  {author} {\bibinfo {author} {\bibfnamefont {KR}~\bibnamefont
  {S{\"u}tterlin}}, \bibinfo {author} {\bibfnamefont {A}~\bibnamefont
  {Wysocki}}, \bibinfo {author} {\bibfnamefont {A.~V.}\ \bibnamefont {Ivlev}},
  \bibinfo {author} {\bibfnamefont {C}~\bibnamefont {R{\"a}th}}, \bibinfo
  {author} {\bibfnamefont {H.~M.}\ \bibnamefont {Thomas}}, \bibinfo {author}
  {\bibfnamefont {M}~\bibnamefont {Rubin-Zuzic}}, \bibinfo {author}
  {\bibfnamefont {W.~J.}\ \bibnamefont {Goedheer}}, \bibinfo {author}
  {\bibfnamefont {V.~E.}\ \bibnamefont {Fortov}}, \bibinfo {author}
  {\bibfnamefont {A.~M.}\ \bibnamefont {Lipaev}}, \bibinfo {author}
  {\bibfnamefont {V.~I.}\ \bibnamefont {Molotkov}}, \bibinfo {author}
  {\bibfnamefont {O.~F.}\ \bibnamefont {Petrov}}, \bibinfo {author}
  {\bibfnamefont {G.~E.}\ \bibnamefont {Morfill}}, \ and\ \bibinfo {author}
  {\bibfnamefont {H.}~\bibnamefont {L\"owen}},\ }\bibfield  {title} {\enquote
  {\bibinfo {title} {Dynamics of lane formation in driven binary complex
  plasmas},}\ }\href@noop {} {\bibfield  {journal} {\bibinfo  {journal} {Phys.
  Rev. Lett.}\ }\textbf {\bibinfo {volume} {102}},\ \bibinfo {pages} {085003}
  (\bibinfo {year} {2009})}\BibitemShut {NoStop}%
\bibitem [{\citenamefont {Sarma}\ \emph {et~al.}(2020)\citenamefont {Sarma},
  \citenamefont {Baruah},\ and\ \citenamefont {Ganesh}}]{SBG2020}%
  \BibitemOpen
  \bibfield  {author} {\bibinfo {author} {\bibfnamefont {Upasha}\ \bibnamefont
  {Sarma}}, \bibinfo {author} {\bibfnamefont {Swati}\ \bibnamefont {Baruah}}, \
  and\ \bibinfo {author} {\bibfnamefont {R}~\bibnamefont {Ganesh}},\ }\bibfield
   {title} {\enquote {\bibinfo {title} {Lane formation in driven pair-ion
  plasmas},}\ }\href@noop {} {\bibfield  {journal} {\bibinfo  {journal} {Phys.
  Plasmas}\ }\textbf {\bibinfo {volume} {27}},\ \bibinfo {pages} {012106}
  (\bibinfo {year} {2020})}\BibitemShut {NoStop}%
\bibitem [{\citenamefont {Shiwakoti}\ \emph {et~al.}(2019)\citenamefont
  {Shiwakoti}, \citenamefont {Shi},\ and\ \citenamefont {Ye}}]{SSY2019}%
  \BibitemOpen
  \bibfield  {author} {\bibinfo {author} {\bibfnamefont {Nirajan}\ \bibnamefont
  {Shiwakoti}}, \bibinfo {author} {\bibfnamefont {Xiaomeng}\ \bibnamefont
  {Shi}}, \ and\ \bibinfo {author} {\bibfnamefont {Zhirui}\ \bibnamefont
  {Ye}},\ }\bibfield  {title} {\enquote {\bibinfo {title} {A review on the
  performance of an obstacle near an exit on pedestrian crowd evacuation},}\
  }\href@noop {} {\bibfield  {journal} {\bibinfo  {journal} {Safety Sci.}\
  }\textbf {\bibinfo {volume} {113}},\ \bibinfo {pages} {54--67} (\bibinfo
  {year} {2019})}\BibitemShut {NoStop}%
\bibitem [{\citenamefont {Takimoto}\ \emph {et~al.}(2002)\citenamefont
  {Takimoto}, \citenamefont {Tajima},\ and\ \citenamefont
  {Nagatani}}]{TTN2002}%
  \BibitemOpen
  \bibfield  {author} {\bibinfo {author} {\bibfnamefont {Kouhei}\ \bibnamefont
  {Takimoto}}, \bibinfo {author} {\bibfnamefont {Yusuke}\ \bibnamefont
  {Tajima}}, \ and\ \bibinfo {author} {\bibfnamefont {Takashi}\ \bibnamefont
  {Nagatani}},\ }\bibfield  {title} {\enquote {\bibinfo {title} {Effect of
  partition line on jamming transition in pedestrian counter flow},}\
  }\href@noop {} {\bibfield  {journal} {\bibinfo  {journal} {Physica A}\
  }\textbf {\bibinfo {volume} {308}},\ \bibinfo {pages} {460--470} (\bibinfo
  {year} {2002})}\BibitemShut {NoStop}%
\bibitem [{\citenamefont {Helbing}\ \emph
  {et~al.}(2000{\natexlab{b}})\citenamefont {Helbing}, \citenamefont {Farkas},\
  and\ \citenamefont {Vicsek}}]{HFV2000a}%
  \BibitemOpen
  \bibfield  {author} {\bibinfo {author} {\bibfnamefont {Dirk}\ \bibnamefont
  {Helbing}}, \bibinfo {author} {\bibfnamefont {Ill{\'e}s~J}\ \bibnamefont
  {Farkas}}, \ and\ \bibinfo {author} {\bibfnamefont {Tam{\'a}s}\ \bibnamefont
  {Vicsek}},\ }\bibfield  {title} {\enquote {\bibinfo {title} {Freezing by
  heating in a driven mesoscopic system},}\ }\href@noop {} {\bibfield
  {journal} {\bibinfo  {journal} {Phys. Rev. Lett.}\ }\textbf {\bibinfo
  {volume} {84}},\ \bibinfo {pages} {1240} (\bibinfo {year}
  {2000}{\natexlab{b}})}\BibitemShut {NoStop}%
\bibitem [{\citenamefont {Nagatani}(2009)}]{Nagatani2009}%
  \BibitemOpen
  \bibfield  {author} {\bibinfo {author} {\bibfnamefont {Takashi}\ \bibnamefont
  {Nagatani}},\ }\bibfield  {title} {\enquote {\bibinfo {title} {Freezing
  transition in the mean-field approximation model of pedestrian counter
  flow},}\ }\href@noop {} {\bibfield  {journal} {\bibinfo  {journal} {Physica
  A}\ }\textbf {\bibinfo {volume} {388}},\ \bibinfo {pages} {4973--4978}
  (\bibinfo {year} {2009})}\BibitemShut {NoStop}%
\bibitem [{\citenamefont {Feliciani}\ and\ \citenamefont
  {Nishinari}(2016)}]{FN2016}%
  \BibitemOpen
  \bibfield  {author} {\bibinfo {author} {\bibfnamefont {Claudio}\ \bibnamefont
  {Feliciani}}\ and\ \bibinfo {author} {\bibfnamefont {Katsuhiro}\ \bibnamefont
  {Nishinari}},\ }\bibfield  {title} {\enquote {\bibinfo {title} {Empirical
  analysis of the lane formation process in bidirectional pedestrian flow},}\
  }\href@noop {} {\bibfield  {journal} {\bibinfo  {journal} {Phys. Rev. E}\
  }\textbf {\bibinfo {volume} {94}},\ \bibinfo {pages} {032304} (\bibinfo
  {year} {2016})}\BibitemShut {NoStop}%
\bibitem [{lam()}]{lammps}%
  \BibitemOpen
  \href@noop {} {}\bibinfo {howpublished} {See \url{http://lammps.sandia.gov}
  for the code}\BibitemShut {NoStop}%
\bibitem [{\citenamefont {Glanz}\ and\ \citenamefont
  {L{\"o}wen}(2012)}]{GL2012}%
  \BibitemOpen
  \bibfield  {author} {\bibinfo {author} {\bibfnamefont {T}~\bibnamefont
  {Glanz}}\ and\ \bibinfo {author} {\bibfnamefont {H}~\bibnamefont
  {L{\"o}wen}},\ }\bibfield  {title} {\enquote {\bibinfo {title} {The nature of
  the laning transition in two dimensions},}\ }\href@noop {} {\bibfield
  {journal} {\bibinfo  {journal} {J. Phys.: Condens. Matter}\ }\textbf
  {\bibinfo {volume} {24}},\ \bibinfo {pages} {464114} (\bibinfo {year}
  {2012})}\BibitemShut {NoStop}%
\bibitem [{\citenamefont {Ikeda}\ \emph {et~al.}(2012)\citenamefont {Ikeda},
  \citenamefont {Wada},\ and\ \citenamefont {Hayakawa}}]{IWH2012}%
  \BibitemOpen
  \bibfield  {author} {\bibinfo {author} {\bibfnamefont {Masahiro}\
  \bibnamefont {Ikeda}}, \bibinfo {author} {\bibfnamefont {Hirofumi}\
  \bibnamefont {Wada}}, \ and\ \bibinfo {author} {\bibfnamefont {Hisao}\
  \bibnamefont {Hayakawa}},\ }\bibfield  {title} {\enquote {\bibinfo {title}
  {Instabilities and turbulence-like dynamics in an oppositely driven binary
  particle mixture},}\ }\href@noop {} {\bibfield  {journal} {\bibinfo
  {journal} {Europhys. Lett.}\ }\textbf {\bibinfo {volume} {99}},\ \bibinfo
  {pages} {68005} (\bibinfo {year} {2012})}\BibitemShut {NoStop}%
\bibitem [{\citenamefont {Reichhardt}\ \emph {et~al.}(2018)\citenamefont
  {Reichhardt}, \citenamefont {Thibault}, \citenamefont {Papanikolaou},\ and\
  \citenamefont {Reichhardt}}]{RTP+2018}%
  \BibitemOpen
  \bibfield  {author} {\bibinfo {author} {\bibfnamefont {Charles}\ \bibnamefont
  {Reichhardt}}, \bibinfo {author} {\bibfnamefont {Joshua}\ \bibnamefont
  {Thibault}}, \bibinfo {author} {\bibfnamefont {S.}~\bibnamefont
  {Papanikolaou}}, \ and\ \bibinfo {author} {\bibfnamefont {C.~J.~O.}\
  \bibnamefont {Reichhardt}},\ }\bibfield  {title} {\enquote {\bibinfo {title}
  {Laning and clustering transitions in driven binary active matter systems},}\
  }\href@noop {} {\bibfield  {journal} {\bibinfo  {journal} {Phys. Rev. E}\
  }\textbf {\bibinfo {volume} {98}},\ \bibinfo {pages} {022603} (\bibinfo
  {year} {2018})}\BibitemShut {NoStop}%
\bibitem [{\citenamefont {Marbach}\ and\ \citenamefont
  {Bocquet}(2017)}]{MB2017}%
  \BibitemOpen
  \bibfield  {author} {\bibinfo {author} {\bibfnamefont {Sophie}\ \bibnamefont
  {Marbach}}\ and\ \bibinfo {author} {\bibfnamefont {Lyd{\'e}ric}\ \bibnamefont
  {Bocquet}},\ }\bibfield  {title} {\enquote {\bibinfo {title} {Active sieving
  across driven nanopores for tunable selectivity},}\ }\href@noop {} {\bibfield
   {journal} {\bibinfo  {journal} {J. Chem. Phys.}\ }\textbf {\bibinfo {volume}
  {147}},\ \bibinfo {pages} {154701} (\bibinfo {year} {2017})}\BibitemShut
  {NoStop}%
\bibitem [{\citenamefont {Marbach}\ and\ \citenamefont
  {Bocquet}(2019)}]{MB2019}%
  \BibitemOpen
  \bibfield  {author} {\bibinfo {author} {\bibfnamefont {Sophie}\ \bibnamefont
  {Marbach}}\ and\ \bibinfo {author} {\bibfnamefont {Lyderic}\ \bibnamefont
  {Bocquet}},\ }\bibfield  {title} {\enquote {\bibinfo {title} {Osmosis, from
  molecular insights to large-scale applications},}\ }\href@noop {} {\bibfield
  {journal} {\bibinfo  {journal} {Chem. Soc. Rev.}\ }\textbf {\bibinfo {volume}
  {48}},\ \bibinfo {pages} {3102--3144} (\bibinfo {year} {2019})}\BibitemShut
  {NoStop}%
\bibitem [{\citenamefont {Zwanzig}(1990)}]{Zwanzig1990}%
  \BibitemOpen
  \bibfield  {author} {\bibinfo {author} {\bibfnamefont {Robert}\ \bibnamefont
  {Zwanzig}},\ }\bibfield  {title} {\enquote {\bibinfo {title} {Rate processes
  with dynamical disorder},}\ }\href@noop {} {\bibfield  {journal} {\bibinfo
  {journal} {Acc. Chem. Res.}\ }\textbf {\bibinfo {volume} {23}},\ \bibinfo
  {pages} {148--152} (\bibinfo {year} {1990})}\BibitemShut {NoStop}%
\end{thebibliography}

%



\clearpage
\renewcommand{\thesection}{S\arabic{section}}
\renewcommand{\thefigure}{S\arabic{figure}}
\setcounter{figure}{0}
\setcounter{section}{0}

\widetext
\section{Supplementary material}
\stepcounter{section}
\section{\label{ssec_method}\thesection.\quad Molecular dynamics simulations}

In the molecular dynamics simulations, we consider a system in which $N$ pedestrians are walking in a straight corridor, as shown in Fig.~2(a).
The half pedestrians are moving in the $+x$ direction (i.e., $\bm{e}_i$ in Eq.~(1) is defined as the unit vector pointing in the $x$ direction) and the remaining half in the $-x$ direction (i.e., $\bm{e}_i$ is the unit vector pointing in the $-x$ direction).

All the simulations are implemented using the open source code LAMMPS~\cite{lammps}.
We employ the velocity Verlet algorithm to time-integrate the equation of motion given in Eq.~(1).
For the interaction forces, the social force model defined in Eq.~(2) is applied.
However, the original LAMMPS package does not include pairwise interactions corresponding to Eq.~(2).
Therefore, in order to realize the social force model with LAMMPS, we modify the subroutine source codes, which are named \verb|pair_buck.cpp| and \verb|pair_gran_hooke.cpp| 
in the original package.

\begin{figure}[b]
\centering
	\includegraphics[width=0.8\hsize]{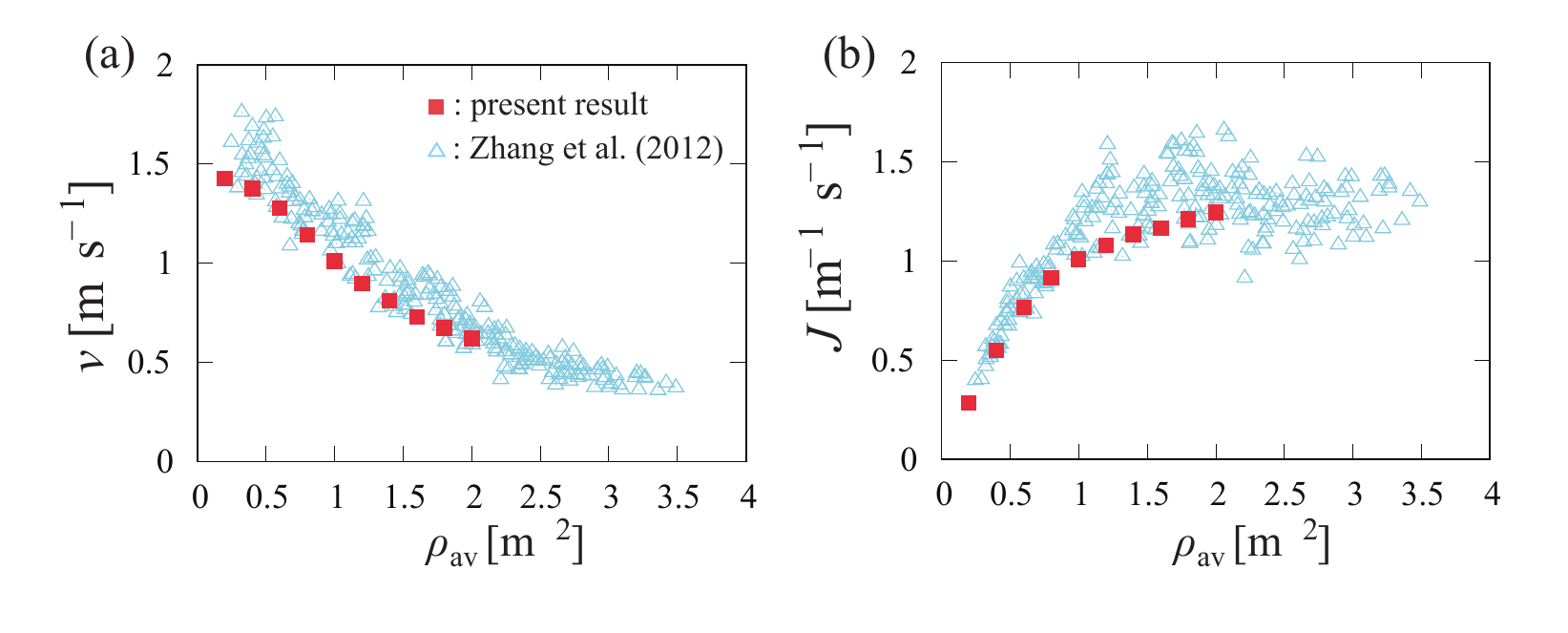}

	\caption{Fundamental diagrams of pedestrian counter flows. (a) Density-velocity and (b) density-flux relations. 
	Here, the flux is defined as $J=\rho_{\mathrm{av}}v$. The width of the corridor in the simulation is $3.6$\,m in accordance with the experiment in Ref.~\cite{ZKS+2012}, whereas it is mainly $8$\,m in the present study.
	\label{fig_fd_validation}}
\end{figure}

The values of model parameters are taken from Ref.~\cite{HFV2000} unless otherwise is stated, 
which is the most widely used parameter set in pedestrian simulations.
In order to validate the model employed and the modified source codes, 
we compare our simulation results with experimental data reported in Ref.~\cite{ZKS+2012}.
The fundamental diagrams, i.e. the velocity-density and flux-density relations 
are given in Fig.~\ref{fig_fd_validation}.
Here the corridor of our simulations has no obstacles and the corridor width is $W=3.6$\,m, matching the situation with the experiment.

\begin{figure}[h]
\centering
        \includegraphics[width=0.8\hsize]{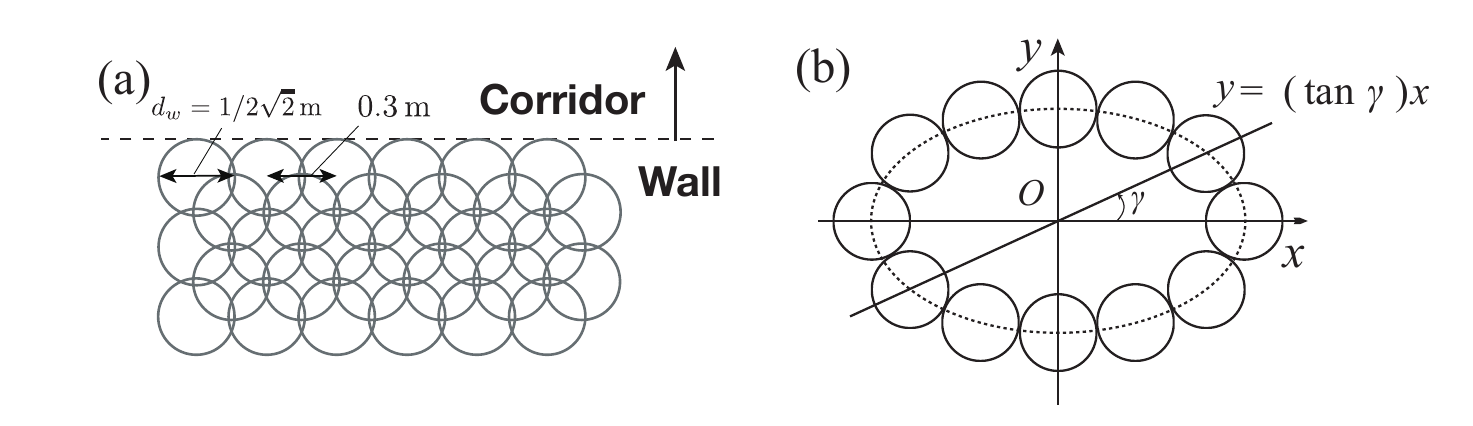}
	\caption{Models for walls and obstacles in terms of fixed particles. (a) straight wall and (b) elliptic obstacle. }
	\label{fig_walls}
\end{figure}

\begin{figure}[h]
     \begin{center}
	\includegraphics[width=0.5\hsize]{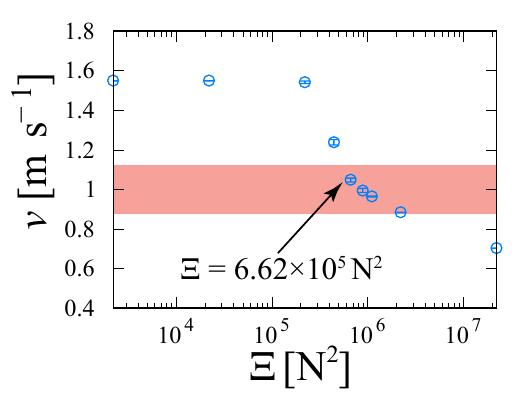}
	\caption{Noise intensity dependence of velocity, under the situation of Fig.~\ref{fig_fd_validation} at $\rho_{\mathrm{av}}=1$\,m$^{-2}$. The red-shaded region represents the range of various experimental data gathered in Ref.~\cite{ZKS+2012}.}
	\label{fig_noise}
	\end{center}
\end{figure}

The corridor walls and obstacles are modeled by groups of fixed particles.
We express each wall in terms of a particle lattice as shown in Fig.~\ref{fig_walls}(a).
On the other hand,
one elliptic obstacle is expressed by $12$ particles located at the points where $x^2/a^2 + y^2/b^2 = 1$ intersects with $y=(\tan \gamma) x$ ($\gamma=n\pi/6,\,(n=1,2,\cdots,12)$),
then it is rotated by angle $\varphi$ about the ellipse's center.
As shown in Fig.~\ref{fig_noise}, the intensity of noise, which is not mentioned in Ref.~\cite{HFV2000},
is determined such that the velocity of pedestrians obtained by the simulations well drops in the
range of experimental values, including Ref.~\cite{ZKS+2012}, with fixing the rest of parameters at the values reported in Ref.~\cite{HFV2000}.

%

%

%
%
\stepcounter{section}
\section{\label{ssec_snap}\thesection.\quad Pedestrian flow patterns}

\begin{figure*}[h]
	\begin{center}
	\includegraphics[width=0.95\hsize]{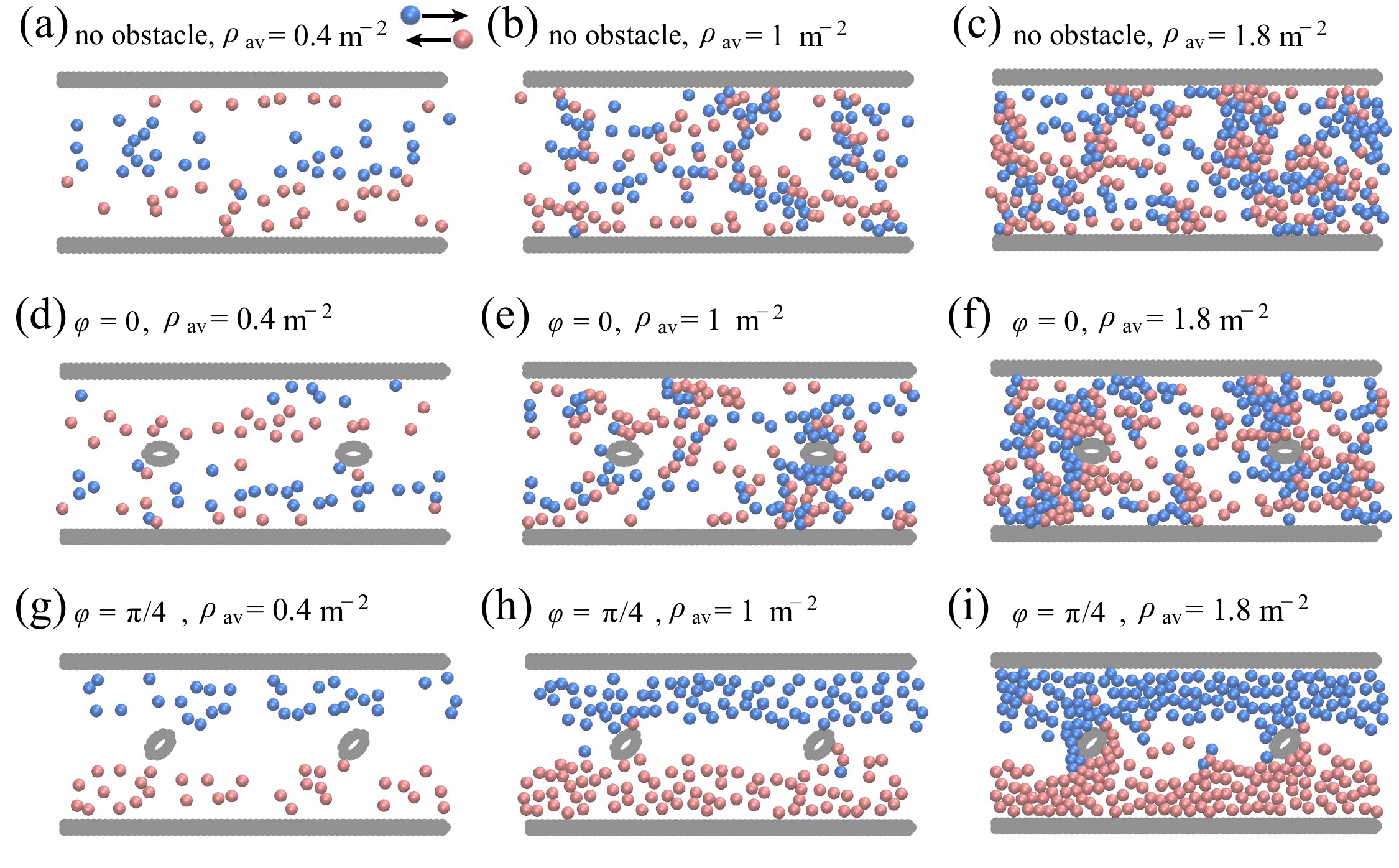}
	\caption{Typical snapshots obtained with the present simulation. Panels (a), (b), and (c) are the snapshots in the case of no obstacles, for 
	$\rho_{\mathrm{av}}=0.4$, $1.0$, and $1.8$\,m$^{-2}$, respectively. The blue (red) particles represent the pedestrians walking in the right (left).
	Panels (d)-(f) are the snapshots in the case of symmetric obstacles ($\varphi =0$),
	and panels (g)-(i) are those in the case of asymmetric obstacles ($\varphi =\pi/4$), for the same values of densities as in panels (a)-(c).}
	\label{fig_snapshots}
	\end{center}
\end{figure*}

\begin{figure*}[h]
	\begin{center}
	\includegraphics[width=0.6\hsize]{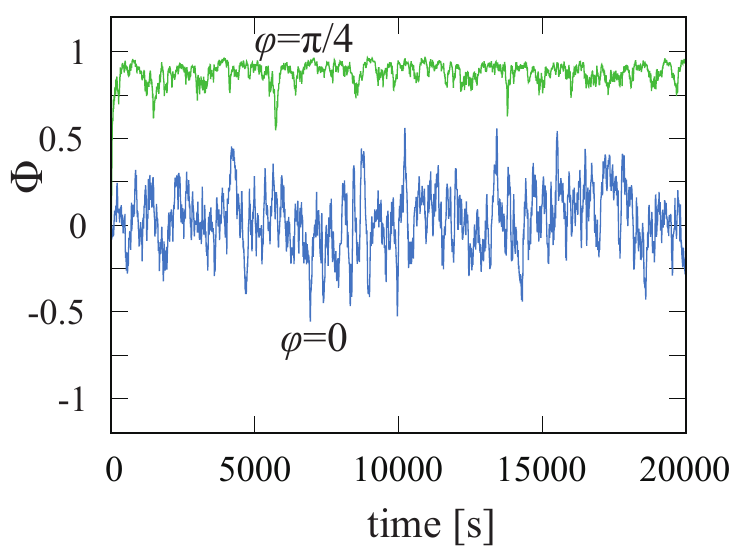}
	\caption{Stability of lanes. (a) time evolution of the order parameter $\Phi$ in the cases of 
	 obstacles with $\varphi=0$ and $\pi/4$, for high density ($\rho_{\mathrm{av}}=1.8$\,m$^{-2}$). }
	\label{fig_instability}
	\end{center}
\end{figure*}

We show typical snapshots obtained with the molecular dynamics simulations
in Fig.~\ref{fig_snapshots} 
for the cases in which there are no obstacle, obstacles of $\varphi=0$, and obstacles of $\pi/4$,.
In the case of $\varphi=\pi/4$ obstacles, pedestrians are separated into two groups moving in the opposite directions,
at any densities, i.e., they always keep left.
We remark in the main text that the observed flow patterns with separation
are stable. This is checked by examining
the time-evolution of the order parameter $\Phi$,
as plotted in Fig.~\ref{fig_instability}.
When $\varphi=\pi/4$ obstacles are placed, the pedestrians do not change their sides while the simulation runs
over $2\times 10^7$ time steps ($2\times 10^4$\, s).
Another point in Fig.~\ref{fig_instability}
is that the fluctuation is larger for the case of $\varphi=0$ than the case of $\varphi=\pi/4$.
This signifies that the pedestrians frequently cross over the median line, to interact with those moving in the other direction,
resulting in active changes of their sides.

%

%
%
\clearpage
\stepcounter{section}
\section{\label{ssec_mod}\thesection.\quad Parameters for the membrane model}

\begin{figure}[h]
	\begin{center}
	\includegraphics[width=0.8\hsize]{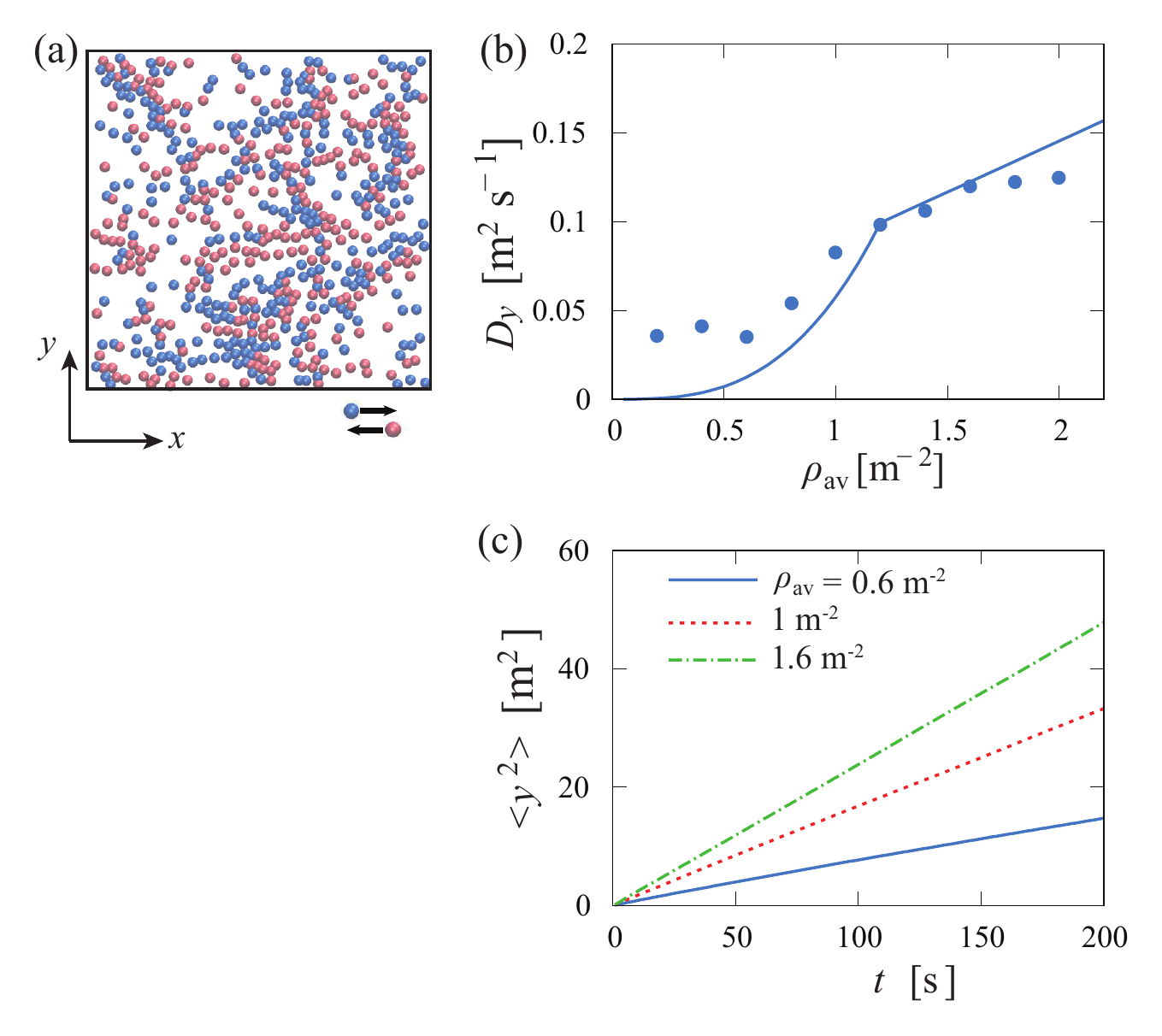}
	\caption{Bulk diffusion coefficient for the present pedestrian counter flows. (a) bulk situation without corridor walls. (b) diffusion coefficient $D_y$ in the $y$ direction
	(which is perpendicular to the desired direction of pedestrians), as a function of average density $\rho_{\mathrm{av}}$.
	The line indicates the corresponding quantity $M_1$ used in the membrane model.
	(c) mean square displacement of particles in the $y$ direction under wall-free bulk conditions, from which the diffusion coefficient $D_y$ is computed.
	}
	\label{fig_diff_coef}
	\end{center}
\end{figure}

First we repeat the developed membrane model equation for separation, given in Eq.~(4) in the main text:
\begin{align}
\dfrac{\mathrm{d}\rho}{\mathrm{d} t}=\dfrac{M_1L}{eS}\left(\dfrac{\rho_\mathrm{av}}{2}-\rho\right)
+\dfrac{M_2L}{eS}\left(\rho_\mathrm{av}-\rho\right),
\label{eq_diff}
\end{align}
%
where the unknown variable $\rho$ is the density of the particles moving in the $+x$ direction in the region  $y>0$ of our corridor.
Under the assumption that the density of all pedestrians
(whichever directions they move to) are uniform in the corridor,
the variable $\rho$ has the relation with the order parameter
as $\Phi=2(\rho/\rho_\mathrm{av}-1/2)$.
The constant $e$ is the thickness of the membrane, and $S=LW/2$ is the area of the region of interest (here the region $y>0$). 
The mobility $M_1$ controls the driving force that mixes the pedestrians such that
the value of $\rho$ approaches $\rho_\mathrm{av}$,
and the mobility $M_2$ corresponds to the driving force that separates the
pedestrians such that $\rho$ approaches $\rho_\mathrm{av}$.
In this study, we use the following models for these mobilities:
\begin{align}
&M_1=
\begin{cases}
\beta_1  L_p \rho_{\mathrm{av}}^{\alpha_1},\quad \rho_{\mathrm{av}}\le \rho_c,\\
 \beta_2 L_p  (\rho_{\mathrm{av}}-\rho_c)^{\alpha_2}+\beta_1 L_p \rho_c^{\alpha_1},\quad \rho_{\mathrm{av}}>\rho_c,
  \end{cases}\label{eq_m1}\\
&M_2= c    L_p^{-1},\label{eq_m2}
\end{align}
where $\alpha_i$ and $\beta_i$ are constants and $c$ is a parameter dependent
on the shape of the obstacles constituting the separation membrane.
Here, the driving force for mixing is assumed to increase with density.
This is based on the fact that the increase in density augments the opportunity to move 
in the direction perpendicular to the traveling direction, i.e., the $y$-direction, avoiding collisions with other pedestrians,
and hence the mean displacement in the $y$-direction increases.
We quantitatively confirm this by means of simulations under wall-free bulk conditions, to calculate the diffusion coefficient from measured mean square displacement in the $y$-direction.
We show  in Fig.~\ref{fig_diff_coef}(a) a snapshot of bulk simulation, and plot in (b) 
the diffusion coefficient in the $y$-direction $D_y$ as a function of the mean density $\rho_{\mathrm{av}}$.
The time dependent of the mean square displacement in the $y$, from which the diffusion coefficient $D_y$ is also shown in (c).

\begin{figure}[h]
	\begin{center}
	\includegraphics[width=0.6\hsize]{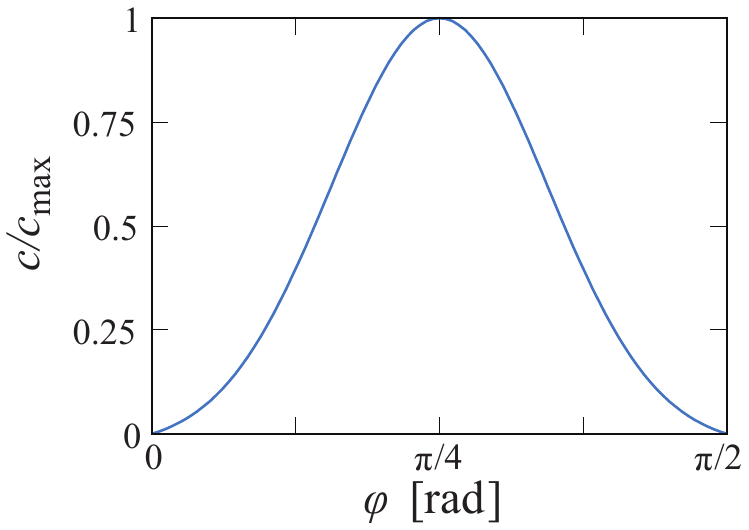}
	\caption{Parameter $c$ included in the filtering membrane model, as a function of the angle of obstacles $\varphi$. }
	\label{fig_cpara}
	\end{center}
\end{figure}

With choosing as $\rho_c=1.2$\,m$^{-2}$ 
and with setting the model parameters as $\alpha_1=3$, $\alpha_2=1$, $\beta_i=1.15\times 10^{-2}$\,m$^{2\alpha_i+1}$/s, we obtain the results shown in Fig.~3 in the main text.
We also compare the value of $M_1$ obtained with these parameters with the diffusion coefficient $D_y$.
In all the model predictions, a common angle dependence of the parameter $c$, which reflects the geometrical details of the obstacles,
is used, which is plotted in Fig.~\ref{fig_cpara} with the normalizing factor $c_{\mathrm{max}}=10$\,m$^3$/s.

%

%
%
\clearpage
\stepcounter{section}
\section{\label{ssec_results}\thesection.\quad Supplemental results}

\begin{figure}[h]
	\begin{center}
	\includegraphics[width=0.55\hsize]{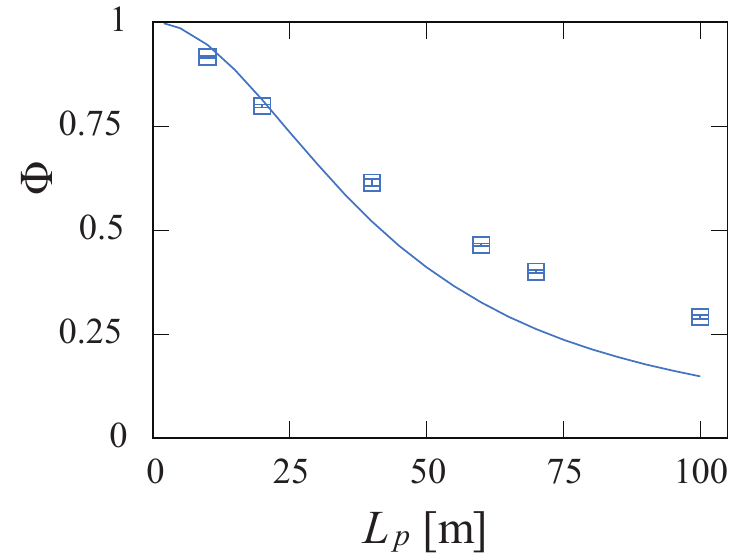}
	\caption{Order parameter $\Phi$ as a function of the interval $L_p$ between obstacles.
	The symbols indicate the simulation results, and the line indicates the predictions of the model given by Eq. (4).}
	\label{fig_phi_lp}
	\end{center}
\end{figure}

Here we provide some supplemental results examining the effect of some parameters.
First we investigate the extended comparison with the membrane model.
We plot the order parameter $\Phi$ as a function of $L_p$ in Fig.~\ref{fig_phi_lp}.
The average density is $\rho_{\mathrm{av}} = 1$\,m$^{-2}$, the
angle of obstacles is $\varphi=\pi/4$, and the corridor width is $W=8$\,m.
The wider $L_p$ becomes, the more opportunities the pedestrians have,
to mix with those passing in the other side of the corridor.
Also, a wide interval $L_p$ reduces the chance for the pedestrians to interact with the
obstacles that yield the driving force for separation.
Our simulation results thus show clear decrease in $\Phi$ with increasing $L_p$.
In our separation membrane model, this effect is accounted for 
by means of the factor $L_p$ in the expressions for $M_1$ and $M_2$, as in Eqs.~\eqref{eq_m1} and \eqref{eq_m2}.
With the factor, the model prediction of the model well reproduces the molecular dynamics results as shown in Fig.~\ref{fig_phi_lp}.

\begin{figure}[h]
	\begin{center}
	\includegraphics[width=1\hsize]{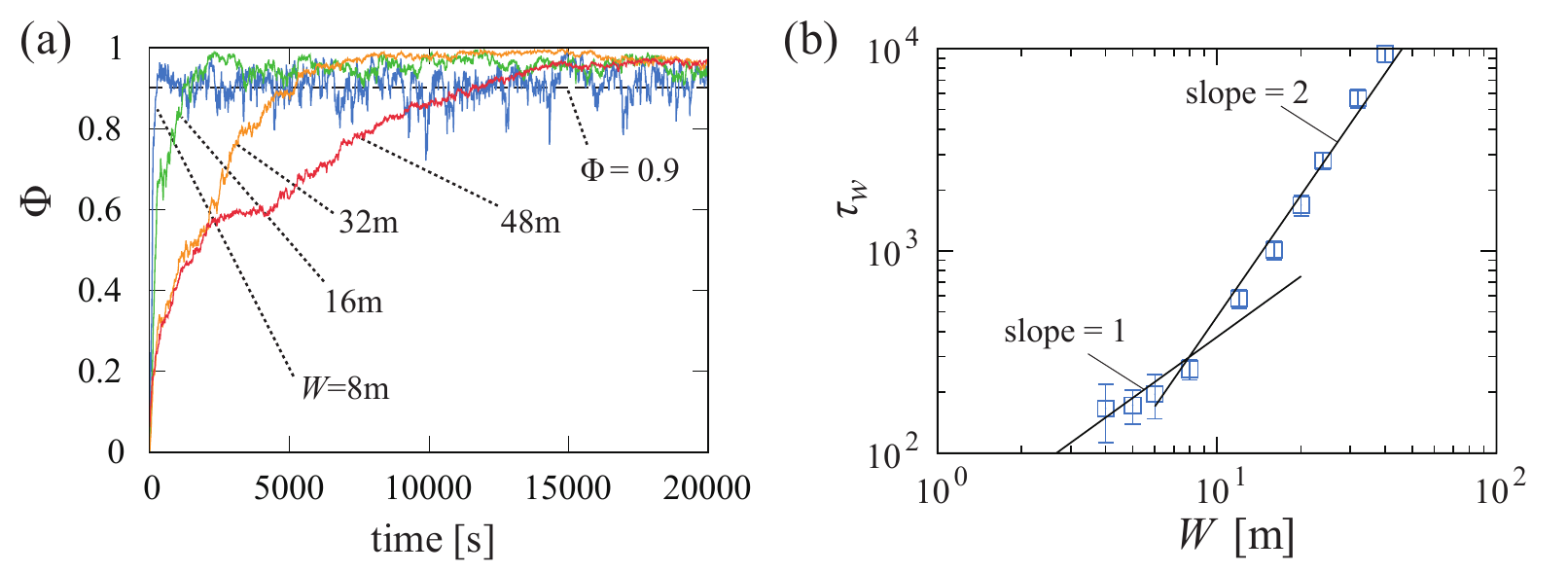}
	\caption{Transient behaviors. (a) time evolution of the order parameter $\Phi$ for various values of the corridor width $W$, in the case of $\rho_{\mathrm{av}}=1$\,m$^{-2}$ 
	and $\varphi=\pi/4$. The dashed line indicates $\Phi=0.9$. The time required to reach $\Phi=0.9$, denoted by $\tau_w$, is plotted in panel (b), as a function of $W$. 
	The lines having slopes equal to unity and two are shown as guides in panel (b).}
	\label{fig_width}
	\end{center}
\end{figure}

We next show in Fig.~\ref{fig_width}(a) the time evolution of $\Phi$ obtained from simulations,
starting from initial conditions in which pedestrians are randomly distributed ($\Phi\sim 0$).
The values of $\Phi$ are plotted for various $W$, fixing the obstacle angle and the average densities
at $\varphi=\pi/4$ and $\rho_{\mathrm{av}}=1$\,m$^{-2}$, respectively.
The systems show $\Phi\sim 1$ after reaching steady states, even for large values of $W$
(up to $W=128$\,m, though not shown in the figure.)
Then we examine the time taken to reach the steady states.
In order to make a quantitative discussion, we define the relaxation time $\tau_w$,
the time interval taken from the initial state to reach $\Phi=0.9$.
As we confirmed in Sec.~\ref{ssec_mod}, the long-time behavior of the pedestrians' displacement in the $y$-direction exhibits diffusion dynamics.
Thus we expect $W^2\sim D_y \tau_w$  for the range $W\gg e$, i.e.
for sufficiently wide corridors, and indeed 
this trend is observed in Fig.~\ref{fig_width}(b).
On the other hand, our membrane model
assumes that the density in each side of the corridor varies uniformely.
In other words, we simplify the model by omitting the diffusion dynamics.
The solution of the model is then obtained as follows:
\begin{equation}
\rho=\rho_0+\rho_1\exp(-M_1Lt/eS)+\rho_2\exp(-M_2Lt/eS),
\end{equation}
where $\rho_1$ and $\rho_2$ are constants and $\rho_0=\rho_{\mathrm{av}}(M_1+2M_2)/(2M_1+2M_2)$ is the stationary solution of the model.
Here $M_2\gg M_1$, because the steady solution corresponds the situation of $\Phi\to 1$, i.e., 
the pedestrians keep left and thus $\rho(t\to \infty)=\rho_0=\rho_{\mathrm{av}}$.
Taking into account this with assuming $\rho_w$ as the value of the density corresponding $\Phi=0.9$, we find
\begin{equation}
\tau_w=-\dfrac{eW}{2M_1}\ln\left(\dfrac{\rho_w-\rho_0}{\rho_1}\right).
\end{equation}
This result implies $\tau_w\sim W$.
This analytical result is consistent with
the behavior observed for relatively narrow corridor ($W\le 8$\,m) in Fig.~\ref{fig_width}(b).
%

\begin{figure}[h]
	\begin{center}
	\includegraphics[width=1.0\hsize]{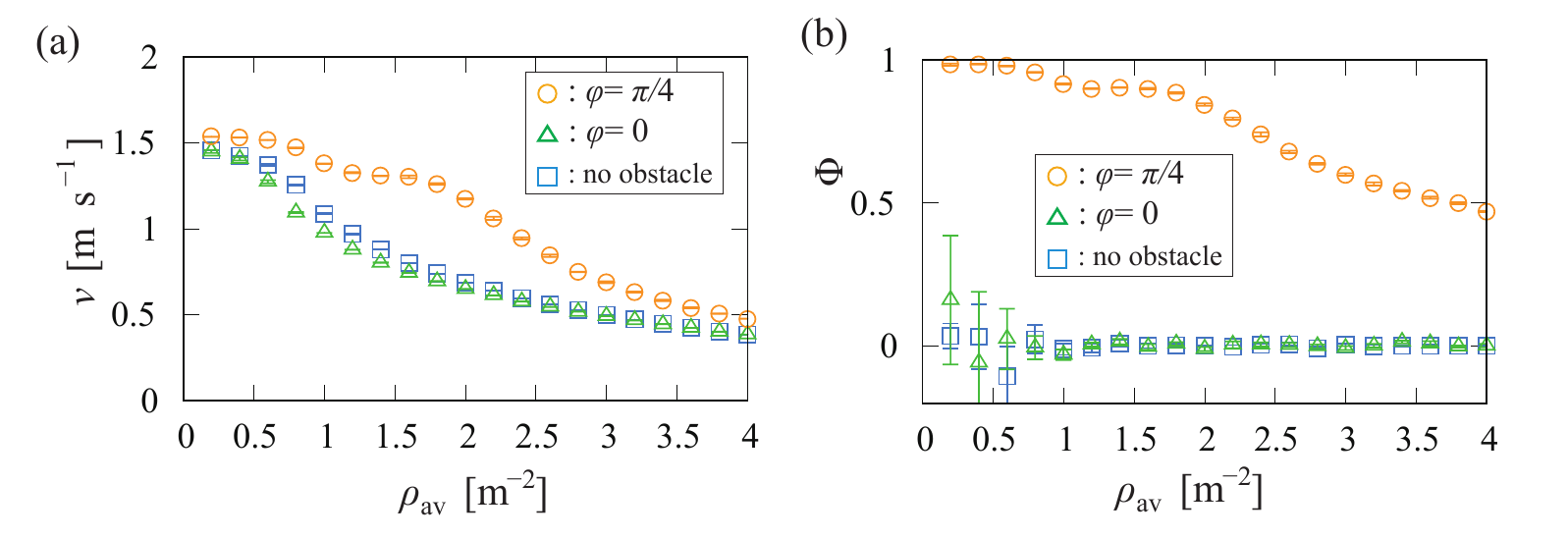}
	\caption{
	(a) Mean velocity $v$ and (b) order parameter $\phi$ versus average density $\rho_{\mathrm{av}}$. The physical and geometrical parameters are the same as those in Fig. 2 of the main text. }
	\label{fig_high_rho}
	\end{center}
\end{figure}

\begin{figure}[h]
	\begin{center}
	\includegraphics[width=0.6\hsize]{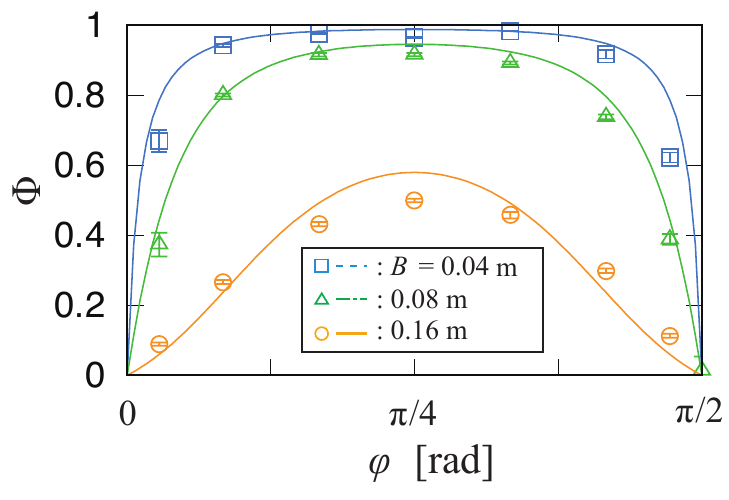}
	\caption{
	Order parameter $\phi$ as a function of angle $\varphi$ in the case of $\rho_{\mathrm{av}}=1$\,m$^{-2}$, for various values of $B$
		appearing in the social force model given in Eq. (2) of the main text.
		 The line indicates the predictions of the model given by Eq. (4), with adjusting the average density.}
	\label{fig_screen}
	\end{center}
\end{figure}

In Fig.~\ref{fig_high_rho}, we show the results for higher average density
than that investigated in the main text:
the mean velocity $v$ in (a) 
and the order parameter $\Phi$ in (b).
The parameter set is the same as Figs. 2 and 3 of the main text.
The separation effect of $\Phi>0.5$ 
is observed up to relatively high density $\rho_{\mathrm{av}}\sim 4$\,m$^{-2}$,
although the effect tends to decay as $\rho_{\mathrm{av}}$ increases.
Accordingly the enhancement in velocity is obtained in this range of density. 
Next we show the effect of the screening parameter $B$ appearing in the 
social force model given in Eq. (2).
Precisely the order parameter $\Phi$ versus the angle $\varphi$ is shown in Fig.~\ref{fig_screen} for different values of $B$,
in the case of $\rho_{\mathrm{av}}=1$\,m$^{-2}$.
The value of $B$ used thoroughly in the main text is $0.08$\,m,
which is the standard value used in the literature.
Whereas the small value $B=0.04$\,m shows the relatively weak impact on the results
with slight enhancement of the separation effect, 
the value of $B$ higher than the standard one, $B=0.16$\,m, weaken the separation effect giving
smaller values of $\Phi$. This is because higher values of $B$ result in wider interaction ranges,
and the size of the particles is thus larger.
This variation of the particle size changes the effective average density.
Indeed the model prediction matches the simulation results with adjusting the average density,
$\rho_{\mathrm{av}}\to 0.6$\,m$^{-2}$ for the case of $B=0.04$\,m and
$\rho_{\mathrm{av}}\to 4.5$\,m$^{-2}$ for the case of $B=0.16$\,m.

%

\end{document}